\documentclass[letterpaper,11pt]{article}
\pdfoutput=1 

\usepackage{jheppub} 

\usepackage{graphicx}
\usepackage{epstopdf}
\usepackage{amsmath, amssymb}
\usepackage{tikz}
\usepackage{verbatim}
\usepackage{caption}
\usepackage{subcaption}
\usepackage{float}
\usepackage{mdframed}
\usepackage{ytableau}
\ytableausetup{smalltableaux, aligntableaux = center}
\def\fund{{\small{\ydiagram{1}}}}
\def\sym{{\small{\ydiagram{2}}}}
\def\antisym{{\small{\ydiagram{1,1}}}}

\newcommand{\be}{\begin{eqnarray}}
\newcommand{\ee}{\end{eqnarray}}

\newcommand{\bn}{\begin{enumerate}}
\newcommand{\en}{\end{enumerate}}

\def\CH{{\cal H}}
\def\CI{{\cal I}}

\def\CN{{\cal N}}
\def\CO{{\cal O}}

\def\a{\alpha}
\def\b{\beta}

\def\w{\omega}

\def\D{\Delta}

\def\Tr{{\rm Tr}}
\def\tr{{\rm tr}}

\usetikzlibrary{arrows,arrows.meta,shapes,shapes.misc,patterns,decorations.pathmorphing,decorations.pathreplacing,positioning,chains,fit,bending,decorations.markings,intersections}

\begin{document}

\title{Supersymmetric Cardy Formula and the Weak Gravity Conjecture in AdS/CFT}

\author[a]{Minseok Cho,}
\author[b]{Sunjin Choi,}
\author[a]{Ki-Hong Lee}
\author[a]{and Jaewon Song}

\affiliation[a]{Department of Physics, Korea Advanced Institute of Science and Technology\\
291 Daehak-ro, Yuseong-gu, Daejeon 34141, Republic of Korea.}

\affiliation[b]{School of Physics, Korea Institute for Advanced Study\\
85 Hoegiro, Dongdaemun-gu, Seoul 02455, Republic of Korea.}

\emailAdd{cms1308@kaist.ac.kr}
\emailAdd{sunjinchoi@kias.re.kr}
\emailAdd{khlee11812@kaist.ac.kr}
\emailAdd{jaewon.song@kaist.ac.kr}

\preprint{KIAS-P23032}

\abstract
{
The Weak Gravity Conjecture (WGC) in anti-de Sitter spacetime (AdS) asserts the existence of an operator in the boundary conformal field theory (CFT) whose scaling dimension-to-charge ratio satisfies a certain upper bound. This bound is specified by the ratio of the conformal central charge $c$ and the flavor central charge $k_F$.
We propose a modified bound in AdS$_5$/CFT$_4$, determined by a combination of two central charges $3c-2a$ instead of $c$. This combination arises in the Cardy-like limit of the 4d superconformal index, which captures the Bekenstein-Hawking entropy of large BPS black holes in AdS$_5$. Using the new bound, we find that certain superconformal field theories (SCFTs) that are previously thought to violate the AdS WGC, including SQCDs in the conformal window, do satisfy the WGC.
We check this version of the WGC against all possible superconformal gauge theories with $SU(N)$ gauge group admitting a large $N$ limit when the superpotential is absent.
We conjecture the modified version of the WGC is a generic property of any 4d SCFT, regardless of the existence of a weakly coupled gravity dual or a large $N$ limit. 
}


\maketitle

\section{Introduction}

The AdS/CFT correspondence \cite{Maldacena:1997re, Gubser:1998bc, Witten:1998qj} is an exact equivalence between a quantum gravity in $d+1$-dimensional anti-de Sitter (AdS) spacetime and a $d$-dimensional conformal field theory (CFT). Newton's constant $G_N$ in AdS gravity is proportional to the inverse of central charge $c$ (which is defined as a coefficient of the singular OPE of the stress-energy tensors) of the dual CFT
\begin{align}
    \begin{split}
        G_N\propto\frac{1}{c}\,.
    \end{split}
\end{align}
Thus, a CFT dual to a weakly coupled gravity should include a large number of degrees of freedom, which typically corresponds to 't Hooft's large $N$ limit \cite{tHooft:1973alw}. This is far from a sufficient condition for a CFT to be dual to a weakly coupled Einstein gravity. There are more conditions to be met, such as the gap in the higher-spin operator spectrum and the sparseness of the spectrum \cite{Heemskerk:2009pn, El-Showk:2011yvt}. 
However, the AdS/CFT correspondence does not demand that the AdS gravity be a weakly coupled Einstein gravity. It implies that any CFT is equivalent to a theory of quantum gravity in AdS, which may be strongly coupled or contain light higher-spin modes, and so on. It is interesting to ask whether there exist universal properties in generic gravitational theories in AdS that may be very different from a weakly coupled Einstein gravity. 

To this end, we study the Weak Gravity Conjecture in the context of AdS/CFT. The Weak Gravity Conjecture (WGC) asserts that in any consistent theory of quantum gravity, there has to be a light-charged particle \cite{Arkani-Hamed:2006emk}. It arises from demanding that any extremal black hole must decay away so that no remnants remain, which can cause a Landau pole problem. This gives a universal condition to be satisfied for any low-energy effective theory that can be consistently coupled to gravity. See \cite{Harlow:2022gzl} and the reference therein. 
The WGC in the CFT \cite{Nakayama:2015hga} can be translated into the statement that there should be an operator whose ratio of the scaling dimension $\Delta$ to the charge $q$ is bounded above by the ratio of the central charge $c$ (which determines the gravitational strength in the bulk) to the flavor central charge $k_F$ (which characterizes the strength of the gauge interaction in the bulk) as
\begin{align} \label{eq:NNWGC}
    \frac{\Delta^2}{q^2} \le \frac{12 c}{k_F}\ .
\end{align}
This version of the WGC has been further tested against a large set of 4d $\CN=1$ superconformal gauge theories, and it was found that the WGC holds even for the highly-exotic theories with a dense spectrum of low-lying operators \cite{Agarwal:2019crm, Agarwal:2020pol}. These results strongly indicate that the WGC holds beyond the weakly coupled Einstein gravity. 

However, it was already found that there exist counterexamples of the WGC \cite{Nakayama:2015hga} in the CFT setup. The 4d supersymmetric quantum chromodynamics (SQCD) with $SU(N)$ gauge group and $N_f$ flavors in the conformal window
\begin{align}
\begin{split}
    \frac{3}{2}N <  N_f < 3N\,
    \end{split}
\end{align}
fails to satisfy the WGC when $N_f/N\gtrsim 2.1$. More precisely, no BPS states satisfy the relation \eqref{eq:NNWGC}. 
This failure can be interpreted in several ways. It may imply that the WGC breaks down at a highly quantum, stringy regime. Or it may mean that there exist light-charged states in the non-BPS sector. However, it is not obvious how it fails, given that we have already observed that the WGC even holds for the highly-exotic theories \cite{Agarwal:2019crm, Agarwal:2020pol}. Hence, a natural question to ask is whether there exists a version of the WGC that holds universally in 
\textit{every} CFT.

One hint comes from the WGC in AdS$_3$/CFT$_2$. In two-dimensional CFTs, it was found that modular invariance of the partition function, which is a universal property of 2d CFTs, was enough to prove the existence of a light-charged state, thereby proving the WGC \cite{Montero:2016tif, Benjamin:2016fhe, Bae:2018qym}. Motivated by this, we look for a similar universal property in 4d. One of the consequences of 2d modular invariance is the celebrated Cardy's formula for the asymptotic free energy (or density of states) at high-energy \cite{Cardy:1986ie}. 
It was recently found that the Cardy-like limit of the 4d superconformal indices \cite{Choi:2018hmj,Kim:2019yrz,Cabo-Bizet:2019osg,Amariti:2019mgp,Cassani:2021fyv} exhibits similar universality. 

The superconformal index of 4d SCFT is defined as
\begin{align}
   \begin{split}
       \mathcal{I}=\Tr_{\CH}\left[ e^{\pi i R}e^{-\omega_1(J_1+R/2)}e^{-\omega_2(J_2+R/2)}\right]\,,
   \end{split}
\end{align}
where the trace is taken over the states satisfying certain BPS conditions, and $J_{1,2}$, $R$ are the Lorentz spins, $R$-charge respectively. 
For this index, the chemical potential $\omega \equiv \frac{\omega_1+\omega_2}{2}$ for the angular momentum $J\equiv \frac{J_{1}+J_2}{2}$ plays the analogous role of the inverse temperature. In the Cardy-like limit, we take $\omega_{1,2}$ to be small, which is an analog of the high-temperature limit. 
In the Cardy-like limit $\omega_{1,2}\rightarrow 0$, the (log of) index behaves universally as
\begin{align} \label{eq:CardyF}
   \begin{split}
       \text{Re}(\text{log}[\CI])\sim(3c-2a)\frac{16\pi^3 \text{Im}(\w_1\w_2)}{27|\w_1|^2|\w_2|^2} \ ,
    \end{split}
\end{align}
which is completely determined via a combination of central charges $a$ and $c$. 
Now, using this asymptotic form of the free energy, one can obtain the entropy $S(J)$ in the microcanonical ensemble at large angular momentum $J$ by the inverse Laplace transformation as follows \cite{Kim:2019yrz}:
\begin{align}\label{eq:Cardy}
    \begin{split}
        S(J)\sim(3c-2a)^{1/3}\cdot J^{2/3} + \CO(J^{1/3})\, \ .
    \end{split}
\end{align}
This formula successfully reproduces the entropy of large BPS black holes in AdS$_5$ \cite{Gutowski:2004ez, Gutowski:2004yv, Chong:2005hr, Kunduri:2006ek}. We note that the entropy scales as $3c-2a$ instead of $c$.

Motivated by the 4d $\CN=1$ Cardy formula \eqref{eq:Cardy}, we suggest a modified version of the WGC for 4d SCFTs:
\begin{mdframed}
    For any 4d $\CN=1$ SCFT with a $U(1)$ \emph{non-R} flavor symmetry $F$, there exists an operator $\CO$ whose scaling dimension $\Delta$ and the charge $q$ satisfy
\begin{align}
    \begin{split}
    \left.\frac{\Delta^2}{q^2}\right|_{\CO}\leq \frac{12(3c-2a)}{k_F
    }\,,
    \end{split}
\end{align} 
where $k_F$ is the flavor central charge $k_F = - 3 \tr\,R F F$.
\end{mdframed}
We simply replaced the central charge $c$ with the combination $3c-2a$ in the WGC proposed by \cite{Nakayama:2015hga}. 
For ordinary holographic theories, the difference between central charges vanishes as we take the large $N$ limit so that $3c-2a \simeq c \sim N^2$. However, for non-holographic theories, $3c-2a$ can differ drastically from $c$. For a generic non-holographic 4d SCFT,  not just $c$ but this particular combination of the central charges encodes the universal spectral behavior in the Cardy-like limit. We emphasize that this relation holds only for a flavor symmetry that is \emph{not} an $R$-symmetry. For the $R$-symmetry, the scaling dimension for a chiral operator is simply fixed by its $R$-charge; therefore, it saturates the original bound \eqref{eq:NNWGC}. 

We test our modified conjecture with every superconformal gauge theory (whose gauge group is simple) that admits a large $N$ limit (including the Veneziano-like limits that are not studied in \cite{Agarwal:2020pol}). Among them, we discuss four counterexamples for the AdS/CFT WGC proposed by Nakayama and Nomura (NN-WGC):
\begin{enumerate}
    \item SQCDs in the conformal window 
    \item $SU(N)$ gauge theory with symmetric, anti-symmetric, and fundamental matters 
    \item $SU(N)$ gauge theory with symmetric and fundamental matters
    \item $SO(10)$ gauge theory with symmetric, anti-symmetric, and fundamental matters
\end{enumerate}
We find that these examples do not satisfy the NN-WGC (in the BPS sector) but do satisfy our modified WGC even for a small finite $N$. This result suggests that the modified WCG holds for generic SCFTs even if the theory is far from being holographically dual to a weakly coupled Einstein-like gravity. 

The organization of the rest of the paper is as follows. In Section \ref{sec:cardy}, we introduce the 4d $\CN=1$ Cardy formula, from which the asymptotic entropy of BPS black holes at large angular momenta can be obtained. From this, we motivate the proposal for the modified WGC in AdS$_5$/CFT$_4$ in Section \ref{sec:WGC}.
In Section \ref{sec:ex}, we provide explicit verification that our modified WGC holds for the four counterexamples listed above.
In Section \ref{sec:large_N_gauge}, we perform comprehensive tests of our new conjecture with all possible 4d superconformal $SU(N)$ gauge theories that admit the large $N$ limit with a vanishing superpotential. 
Then we conclude with discussions in Section \ref{sec:discussion}.

\section{Supersymmetric Cardy formula and the Weak Gravity Conjecture}

\subsection{Black hole entropy from supersymmetric Cardy formula}
\label{sec:cardy}

In this subsection, we review the Cardy formula of the superconformal index of 4d $\CN=1$ SCFT \cite{Choi:2018hmj,Kim:2019yrz,Cabo-Bizet:2019osg,Amariti:2019mgp,Cassani:2021fyv}. This Cardy-like formula for the index captures the entropy of large BPS black holes in AdS$_5$ \cite{Gutowski:2004ez, Gutowski:2004yv, Chong:2005hr, Kunduri:2006ek}. The superconformal index \cite{Kinney:2005ej,Romelsberger:2005eg} of the 4d $\mathcal{N}=1$ SCFT is defined as
\begin{align}
    \begin{split}
        \CI=\Tr_{\CH}\left[(-1)^Fe^{-\omega_1(J_1+R/2)}e^{-\omega_2(J_2+R/2)} \prod_I z_I^{f_I} \right]\ ,
    \end{split}
    \label{eq:index}
\end{align}
where the trace is taken over the radially quantized Hilbert space $\CH$ on $S^3$. Here, $F$ is the fermion number operator, $J_{1,2}$ are the Cartan charges of $SO(4)$ rotation symmetry, $R$ is the $U(1)_R$ $R$-charge, and $f_I$'s are the Cartan charges of the flavor symmetry.
The index receives contribution only from the $\frac{1}{4}$-BPS states satisfying the BPS condition
\begin{align}
    \begin{split}
        \Delta-J_1-J_2-\frac{3}{2}R=0\,,
    \end{split}
\end{align}
where $\Delta$ is the scaling dimension. As usual, it is invariant under any continuous deformation of the theory. Also, note that it is not periodic under the shifts $\omega_{1,2}\rightarrow \omega_{1,2}+2\pi i$ since the $R$-charges of the operators in a generic 4d $\CN=1$ SCFT is not integer-quantized. 
Redefining $\omega_1$ as $\omega_1 \to \omega_1 -2\pi i$, the index \eqref{eq:index} can be rewritten as
\begin{align}
    \begin{split}
        \mathcal{I}=\Tr_{\CH}\left[ e^{\pi i R}e^{-\omega_1(J_1+R/2)}e^{-\omega_2(J_2+R/2)} \prod_I z_I^{f_I} \right]\ .
    \end{split}
    \label{eq:modindex}
\end{align}

The Cardy limit of the index \eqref{eq:modindex} was analyzed using the saddle point approximation of its matrix integral formula for the gauge theories \cite{Choi:2018hmj,Kim:2019yrz,Cabo-Bizet:2019osg,Amariti:2019mgp} and the anomaly-based effective field theory analysis on $S^3$ \cite{Choi:2018hmj,Kim:2019yrz,Cassani:2021fyv}.\footnote{In both analysis, vanishing gauge holonomy configuration was assumed, which is a local saddle point of every SCFTs in the Cardy limit, thereby yielding the universal asymptotic formula of the index. In fact, for a large number of examples, it gives the dominant contribution to the index \cite{Kim:2019yrz}.} The asymptotic behavior of \eqref{eq:modindex} in the Cardy limit $\omega_{1,2} \to 0$ is given by
\begin{align}
    \text{log}(\mathcal{I})\sim\Tr\, R^3\frac{\delta^3}{6\omega_1\omega_2}+\Tr\, R\frac{\pi^2\delta}{6\omega_1\omega_2}=\frac{8(5a-3c)}{27\omega_1\omega_2}\delta^3+\frac{8\pi^2(a-c)}{3\omega_1\omega_2}\delta\, ,
    \label{eq:Cardy2}
\end{align}
where $\delta = \frac{\omega_1+\omega_2 -2 \pi i}{2}$, and the flavor fugacities are all turned off for simplicity.
Performing the Legendre transformation in the unrefined case $\omega = \omega_1 = \omega_2$, the asymptotic entropy at large angular momentum $J\equiv \frac{J_1+J_2}{2}$ is given by
\begin{align}
\text{Re}(S)=2^{1/3}3^{1/2}(3c-2a)^{1/3}\pi\cdot J^{2/3}+O(J^{1/3})\,.
\label{eq:Cardyentropy}
\end{align}
Note that this is guaranteed to be positive thanks to the Hofman-Maldacena bound $\frac{1}{2}<\frac{a}{c}<\frac{3}{2}$ for $\CN=1$ SCFTs \cite{Hofman:2008ar}.

In holographic theories, one can further take the limit in which the central charges $a \sim c$ become large. On the dual AdS gravity side, this corresponds to the semi-classical limit where the Newton constant $G_N$ becomes small. Then, \eqref{eq:Cardyentropy} perfectly captures the entropy of the large BPS black holes in AdS$_5$ \cite{Gutowski:2004ez,Gutowski:2004yv,Chong:2005hr,Kunduri:2006ek}. 
The BPS black holes exist well beyond the Cardy limit and it has been shown that the index correctly captures their entropies \cite{Benini:2018ywd, Choi:2021rxi}.
Moreover, in some holographic models, it has been shown that \eqref{eq:Cardyentropy} matches the four-derivative corrected black hole entropy in 5d $\CN=1$ minimal gauged supergravity \cite{Bobev:2021qxx,Bobev:2022bjm,Cassani:2022lrk}.

\subsection{Weak Gravity Conjecture in AdS/CFT}
\label{sec:WGC}
In this subsection, we review the Weak Gravity Conjecture in AdS$_5$, its implication in the dual CFT$_4$, and also propose a modified conjecture that applies to arbitrary 4d $\mathcal{N}=1$ SCFTs. 
The Weak Gravity Conjecture is motivated by the idea that every black hole should decay unless it is forbidden either kinematically or by symmetry \cite{Arkani-Hamed:2006emk}.
Suppose that an extremal Reissner-Nordstr\"om black hole decays by emitting a particle of mass $m$ and charge $q$. After the emission, the black hole should also satisfy the extremal bound as
\begin{align}
    M_{\text{ext}}(Q)-m\geq M_{\text{ext}}(Q-q)\simeq M_{\text{ext}}(Q)-q\frac{\partial M_{\text{ext}}}{\partial Q}\,.
\end{align}
It gives rise to the bound on the mass-to-charge ratio of the emitted particle:
\begin{align}
\frac{m}{q}\leq \frac{\partial M_{\text{ext}}}{\partial Q}\,.
\label{eq:WGCgen}
\end{align}
The Weak Gravity Conjecture demands that such decay should occur. In other words, there should be a light particle, called a super-extremal particle, satisfying the above condition \eqref{eq:WGCgen}. 
For the Einstein-Maxwell theory in 5d flat spacetime, the extremal mass $M_{\text{ext}}\propto Q$, and the upper bound is fixed by the constant $M_{\text{ext}}/Q=\sqrt{3g^2/2G_N}$. This bound can be corrected by the higher derivative terms \cite{Kats:2006xp,Cheung:2018cwt,Hamada:2018dde}. 

In AdS$_5$ spacetime, the derivative $\frac{\partial M_{\text{ext}}}{\partial Q}$ of the extremal Reissner-Nordstr\"om mass is a monotonically increasing function in $Q$ whose minimum is at $Q\rightarrow0$ limit\footnote{This is also a limit in which the black holes become BPS solutions. We are taking the limit where the size of the black hole is still above a certain cutoff scale such that it is still a macroscopic object. One natural candidate cutoff is the Planck scale where the extremal black hole mass $\sim G_N^{-\frac{1}{3}}$ \cite{Nakayama:2015hga}. There might be a larger cutoff scale, for example, as suggested in \cite{Arkani-Hamed:2006emk} by the ``UV brane," which restricts the black hole mass $m\gtrsim G_N^{-3/5}$.} so that the strongest condition of the WGC arises as \cite{Nakayama:2015hga} 
\begin{align}
    \frac{m^2}{q^2}\leq\frac{3g^2}{2G_N}\,.
    \label{eq:adswgc}
\end{align}
Note that \eqref{eq:adswgc} is exactly the same condition as that on the flat spacetime because the $Q\to0$ limit of the AdS black hole is equivalent to the large AdS length scale limit $L\to \infty$, i.e., the flat spacetime limit.

\begin{figure}[h]
\centering
\begin{subfigure}[b]{0.25\linewidth}
\includegraphics[width=\linewidth]{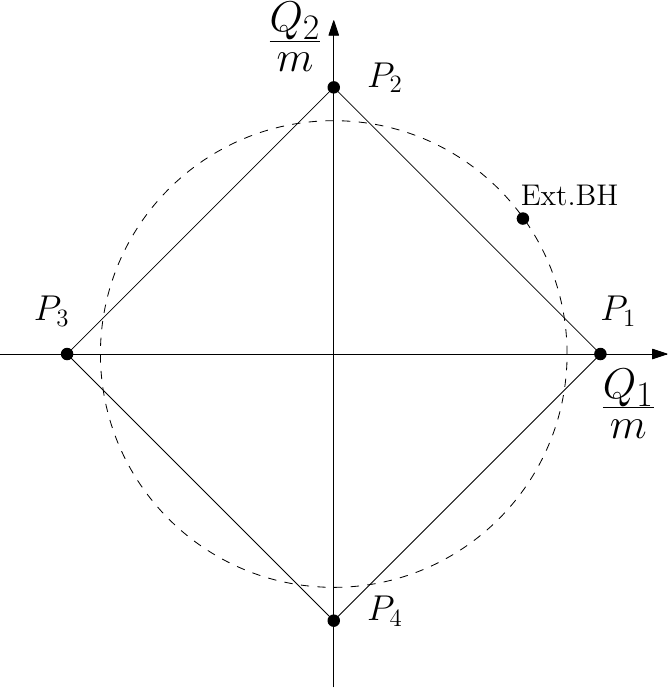}
\subcaption{}
\end{subfigure}
\hspace{15mm}
\begin{subfigure}[b]{0.27\linewidth}
\includegraphics[width=\linewidth]{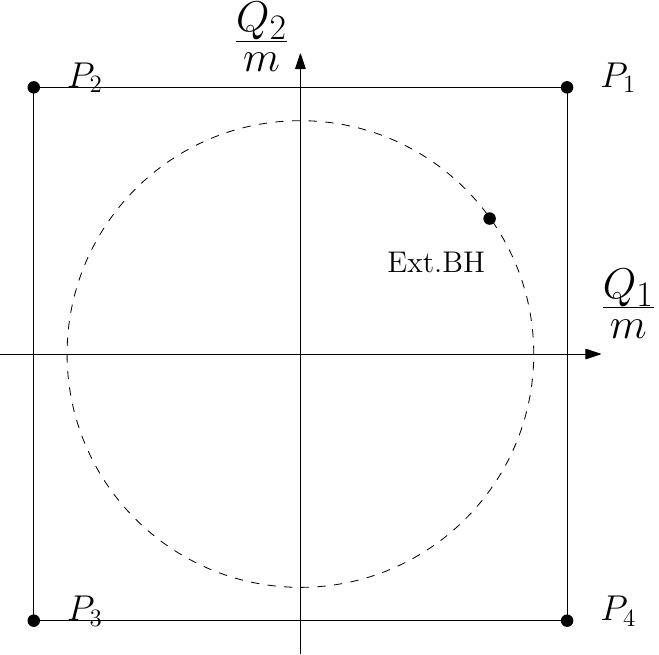}
\subcaption{}
\end{subfigure}
\caption{The charge-to-mass ratio space with two flavor charges, normalized by the ratio of the extremal black holes. An extremal black hole outside the convex hull cannot decay by emitting charged particles. When the convex hull encloses the unit ball, any extremal black holes can decay by emitting particles.}
\label{fig:cvh_gen}
\end{figure}

In a gravity theory with $U(1)^n$ gauge symmetry, the presence of $n$ light particles $P_{i=1,\dots,n}$, where each $P_i$ satisfies \eqref{eq:adswgc} with respect to $i$-th $U(1)$ gauge symmetry, is insufficient for a Reissner-Nordstr\"om black hole to decay. This is because  an extremal Reissner-Nordstr\"om black hole with charges $\vec{Q}$ 
has its mass\footnote{We are considering $|\vec{Q}|\rightarrow 0$ limit where the WGC condition is the strongest. } $M_{\text{ext}}(\vec{Q})\propto|\vec{Q}|$ that is smaller than the sum of the extremal masses of single-charged black holes, i.e.,
\begin{align}
    M_{\text{ext}}(\vec{Q})\leq \sum_{i=1}^n M_{\text{ext}}(Q_i)\ .
\end{align}
In order for a multi-charged black hole to decay, a stronger condition so-called ``convex hull condition" \cite{Cheung:2014vva} should meet. The convex hull condition claims that we should be able to find a set of light particles, which forms a convex hull in the charge-to-mass ratio space that encloses the unit ball representing the set of multi-charged extremal Reissner-Nordstr\"om black holes. Figure \ref{fig:cvh_gen} (a) depicts the case when the convex hull condition is not met even though each particle satisfies the WGC with respect to the individual $U(1)$ symmetries, and Figure \ref{fig:cvh_gen} (b) depicts the case which satisfies the convex hull condition. We see that the particles $P_{1,2,3,4}$ in (b) have much larger charge-to-mass ratios (distance from the origin) than those in (a).

Now we consider what the WGC implies on CFT$_4$ through the AdS/CFT correspondence \cite{Nakayama:2015hga}. The AdS/CFT correspondence says that a state with mass $m$ in an AdS$_5$ corresponds to an operator of scaling dimension $\Delta$ in CFT$_4$ with the relation
\begin{align}
    \Delta\simeq Lm\,,
\end{align}
where $L$ is the AdS length scale. Also, note that the global symmetry in CFT corresponds to the gauge symmetry in dual AdS gravity. The Newton constant $G_N$ and the gauge coupling $g$ in AdS are related to the conformal central charge $c$ and the (non-$R$) flavor central charge $k_F\equiv- 3 \tr\,R F F$ of the dual CFT$_4$ as
\begin{align}
C_T\equiv\frac{40}{\pi^4}c=\frac{40L^3}{\pi^2}{G_N}^{-1}\,,\quad   C_F\equiv \frac{3}{4\pi^4}k_F=\frac{6L}{\pi^2}g^{-2}\,.
\label{eq:corr}
\end{align}
Therefore, the Weak Gravity Conjecture demands that the 4d CFT should contain at least one operator $\mathcal{O}$ whose scaling dimension $\Delta$ and the charge $q$ satisfy the following condition:
\begin{align}
    \left.\frac{\Delta^2}{q^2}\right|_{\mathcal{O}}\leq\frac{9}{40}\frac{C_T}{C_F}=\frac{12c}{k_F}\,.
    \label{eq:CFTWGC1}
\end{align}
When a CFT has more than one $U(1)$ flavor symmetry, the convex hull condition must hold, where the charge-to-mass ratio is translated into a charge-to-dimension ratio in the unit of $\sqrt{\frac{k_F}{12c}}$ from \eqref{eq:CFTWGC1}. 

So far, our discussion is based on a semi-classical analysis of weakly coupled Einstein-Maxwell theory in AdS and its translation to the CFT language. 
There is no obvious reason to expect such a bound to be extended to non-holographic CFTs that are not dual to the weakly coupled Einstein gravity. Indeed, 4d $\CN=1$ SQCDs in the Veneziano-like limit are shown to violate this bound when $N_f/N\gtrsim2.1$ \cite{Nakayama:2015hga}\footnote{More precisely, there is no chiral operator satisfying the WGC bound. This does not rule out the possibility of having non-BPS operators satisfying the bound, but we think this to be highly unlikely. In this paper, whenever we mention the violation of WGC, we mean that such operators do not exist in the chiral ring.}. 
However, it has also been shown that highly exotic (non-holographic) theories with `dense' spectra also obey the WGC \cite{Agarwal:2019crm, Agarwal:2020pol}. Therefore, it is interesting to ask whether there is a universal version of the AdS/CFT WGC that also holds for arbitrary CFTs. In 2d CFT, a super-extremal state can always be found due to the modularity \cite{Benjamin:2016fhe,Montero:2016tif,Bae:2018qym}. The modularity also gives rise to the density of states being proportional to the central charge $c$ at high-energy of 2d CFT \cite{Cardy:1986ie}. In 4d SCFT, the density of states also exhibits similar universality as reviewed in Section \ref{sec:cardy}. It scales as $3c-2a$ rather than $c$ at high energy. Comparing this with \eqref{eq:CFTWGC1}, we suggest a modified WGC bound on the dimension-to-charge ratio of a light operator $\mathcal{O}$ as
\begin{align}
    \left.\frac{\Delta^2}{q^2}\right|_{\mathcal{O}}\leq \frac{12(3c-2a)}{k_F}\,,
    \label{eq:CFTWGC2}
\end{align}
where the coefficient $c$ is replaced by $3c-2a$. The convex hull condition should also be modified with a new unit of charge-to-dimension ratio $\sqrt{\frac{k_F}{12(3c-2a)}}$. Due to the Hofman-Maldacena bound \cite{Hofman:2008ar}, the RHS of \eqref{eq:CFTWGC2} is always positive for an interacting SCFT. The RHS vanishes for a free vector multiplet, but this theory does not have a (non-$R$) flavor symmetry. 
Notice that the new bound is weaker for the CFTs with $c>a$ but stronger for the CFTs with $c<a$. 
In the rest of the paper, we test our proposal extensively against a large set of 4d SCFTs.

Let us emphasize that \eqref{eq:CFTWGC2} only applies to {\it{non}}-$R$ symmetries. Notice that every chiral operator $\mathcal{O}$ satisfies the condition
\begin{align}
\Delta_{\mathcal{O}}=\frac{3}{2}R_{\mathcal{O}}\, \ , 
\label{eq:chir}
\end{align}
so that it saturates the original bound \eqref{eq:CFTWGC1} where $\tfrac{16}{3}c$ replaces $k_F$ for the $R$-symmetry. On the other hand, all chiral operators violate the new bound for the $R$-charge when $c<a$ since the upper bound in \eqref{eq:CFTWGC2} is smaller than the RHS of \eqref{eq:CFTWGC1}. Therefore we only consider the \emph{flavor} symmetries in the rest of this paper.

\section{Examples violating the NN-WGC}
\label{sec:ex}

In this section, we find a number of examples that violate the AdS Weak Gravity Conjecture of Nakayama-Nomura (NN-WGC) \cite{Nakayama:2015hga} but obey the modified WGC \eqref{eq:CFTWGC2}.

\subsection{SQCD in the conformal window}
\label{sec:SQCD}
\paragraph{SU SQCD}
As a first example, let us consider the 4d $\CN=1$ $SU(N)$ SQCD. The IR behavior of SQCD depends on the number of flavors $N_f$, and it flows to an interacting IR SCFT when $N_f$ lies within the conformal window \cite{Seiberg:1994pq}
\begin{align}
    \frac{3}{2}N < N_f < 3N\,.
    \label{eq:conwin}
\end{align}
In this theory, we cannot take a simple large $N$ limit with $N_f$ fixed while staying inside the conformal window. Therefore, we take a so-called Veneziano-like limit, taking large $N$ while the ratio $N_f/N$ is fixed. 
The large $N$ limit of the IR SCFT of SQCD in the conformal window is not expected to be dual to a weakly coupled Einstein gravity in AdS  since the two central charges $a$ and $c$ are not identical in the large $N$ limit. It was found that the $SU(N)$ SQCD violates the NN-WGC when the flavor-to-color ratio $N_f/N$ is larger than a certain value \cite{Nakayama:2015hga}. On the other hand, we will find that our modified WGC is satisfied for arbitrary $N_f/N$ as long as we are within the conformal window.

The $R$-charge at the IR fixed point is determined by the anomaly-free condition for the $R$-symmetry
\begin{align}
    \Tr RGG=T(\textbf{adj})+N_f\,T(\,{\small\fund}\,)(R_Q-1)+N_f\,T(\,{\small\overline{\fund}}\,)(R_{\widetilde{Q}}-1)=0\,,
    \label{eq:anomfreeR}
\end{align}
where the $T(\mathbf{R})$ is the Dynkin index of the representation $\mathbf{R}$, which is given as $T(\textbf{adj})=N$, $T(\,{\small\fund}\,)=T(\,{\small\overline{\fund}}\,)=1/2$. The charge of each field is listed in Table \ref{tab:su_tab}. 
{\renewcommand\arraystretch{1.6}
\begin{table}[h!]
\centering
\begin{tabular}{|c|c||c|c|c|}
\hline
\# & Fields & $SU(N)$ & $U(1)_B$ & $U(1)_R$\\
\hline
$N_f$ & $Q$ & ${\small\fund}$ & $1$ & $1-\frac{N}{N_f}$\\
$N_f$ & $\widetilde{Q}$ & ${\small\overline{\fund}}$ & $-1$ & $1-\frac{N}{N_f}$\\
\hline
\end{tabular}
\caption{The matter contents and their charges in the $SU(N)$ SQCD with $N_f$ matters}
\label{tab:su_tab}
\end{table}}
\\
At the IR fixed point, the central charges $a$, $c$ are computed by anomaly coefficients \cite{Anselmi:1997am} as
\begin{align}
\begin{split}
    a &=\frac{3}{32}\left(3\Tr\,R^3-\Tr\,R\right)=\frac{3}{16}\left(2N^2-3\frac{N^4}{N_f^2}-1\right)\,,\\
    c &=\frac{1}{32}\left(9\Tr\,R^3-5\Tr\,R\right)=\frac{1}{16}\left(7N^2-9\frac{N^4}{N_f^2}-2\right)\,,
    \end{split}
\end{align}
and the flavor central charge for the baryonic $U(1)_B$ is given as
\begin{align}
    k_B & \equiv -3\text{Tr}RBB=6N^2\,.
\end{align}
The corresponding upper bound for the scaling dimension to the baryonic charge ratio of the operator required by the Weak Gravity Conjecture is
\begin{align}
    \begin{split}
        \frac{12(3c-2a)}{k_B}=\frac{9}{8}\left(1-\frac{N^2}{N_f^2}\right)\,.
    \end{split}
    \label{eq:su_bd}
\end{align}
Then, we consider the baryonic operator 
\begin{align}
\epsilon_{a_1\cdots a_N}Q_{I_1}^{a_1}\cdots Q_{I_N}^{a_N}\,,
\end{align}
(or shortly $\epsilon Q^N$,) where the gauge index $a_i$ runs from $1,\dots,N$ and flavor index $I_j$ runs from $1,\dots,N_f$. The $\epsilon_{a_1\cdot a_N}$ is the completely anti-symmetric tensor of $SU(N)$.
It is a chiral operator with the scaling dimension $\Delta=\frac{3}{2}R_{\epsilon Q^N}=\frac{3}{2}NR_Q$ and the baryonic charge $q_B=N$, and their ratio at the IR fixed point is given by
\begin{align}
    \begin{split}
    \left.\frac{\Delta^2}{q_B^2}\right|_{\epsilon Q^N}=\frac{\left(\frac{3}{2}N\left(1-\frac{N}{N_f}\right)\right)^2}{N^2}=\frac{9}{4}\left(1-\frac{N}{N_f}\right)^2\,.
    \end{split}
    \label{eq:su_bar}
\end{align}
Comparing \eqref{eq:su_bd} and \eqref{eq:su_bar}, one can easily see that the baryonic operator always satisfies the modified WGC
\begin{align}
    \begin{split}
        \left.\frac{\Delta^2}{q_B^2}\right|_{\epsilon Q^N}< \frac{12(3c-2a)}{k_B}
    \end{split}
\end{align}
with respect to the $\frac{N}{N_f}$ in the conformal window range $\frac{1}{3}<\frac{N}{N_f} < \frac{2}{3}$. Especially, the baryon is strictly below the modified WGC bound because of the upper bound of the conformal window $N_f<3N$.

\paragraph{SO/Sp SQCD}
Next, we consider the $SO(N)$ SQCD with $N_f$ matters in $SO(N)$ vector representation. For an SQCD to flow to an interacting IR fixed point, the number of flavors should be in the conformal window \cite{Intriligator:1995id}
\begin{align}
    \frac{3}{2}(N-2) < N_f< 3(N-2)\,.
    \label{eq:SOconwin}
\end{align}
Unlike the $SU$ SQCDs, where both fundamental and anti-fundamental matters come in pairs (to cancel gauge anomaly), the $SO$ SQCD has matters in vector representation only. As a result, there is no anomaly-free $U(1)$ flavor symmetry other than $U(1)_R$ symmetry. However, we can still consider the WGC with the non-abelian $SU(N_f)$ flavor symmetry in a similar manner by considering its subgroup $U(1)_B$ symmetry. Consider the most general $U(1)_B$ subgroup of the $SU(N_f)$ flavor symmetry under which each matter chiral multiplet $Q_I$ is charged as
\begin{align}
    \begin{split}
        q_B[Q_I]=q_I\,,\quad I=1,\dots,N_f\,.
    \end{split}
\end{align}
It should satisfy the anomaly-free condition
\begin{align}
    \begin{split}
        \text{Tr}BGG=\sum_{I=1}^{N_f}q_I=0\,.
    \end{split}
    \label{eq:traceless}
\end{align}
Here, we assume the $q_0=\text{max}(|q_1|,\cdots,|q_{N_f}|)\neq0$ so that it is charged by at least one of the matter fields. However, any $U(1)$ subgroup of non-abelian flavor symmetry is traceless and does not mix with the $R$-symmetry at IR \cite{Intriligator:2003jj}. Then the $R$-charge of each matter chiral field $Q_I$ is identically fixed as
\begin{align}
    \begin{split}
        R_{Q_I}=1-\frac{N-2}{N_f}\,,\quad I=1,\cdots,N_f
    \end{split}
\end{align}
by the anomaly-free condition $\Tr RGG=0$, which is the same with \eqref{eq:anomfreeR} without $\widetilde{Q}$ fields. The corresponding IR central charges are
\begin{align}
    \begin{split}
        a & =\frac{3}{32}\left(N(N-1)+N(N-2)-\frac{3N(N-2)^3}{N_f^2}\right)\,,\\
        c & =\frac{1}{32}\left(2N(N-1)+5N(N-2)-\frac{9N(N-2)^3}{N_f^2}\right)\,,
    \end{split}
\end{align}
and the flavor central charge is given as
\begin{align}
        k_B & \equiv -3\text{Tr} RBB =\frac{3N(N-2)}{N_f}\sum_{I=1}^{N_f}q_I^2\,.
\end{align}
Now we consider the operator 
\begin{align}
\delta_{ab}Q_{I_0}^aQ_{I_0}^b
\end{align}
(or shortly $Q_{I_0}^2$) where the gauge index $a,b$ runs from $1,\cdots,N$ and $Q_{I_0}$ is the matter field whose $U(1)_B$ charge is $q_0$ in the absolute value, i.e., $|q_B[Q_{I_0}]|=|q_0|\geq |q_B[Q_I]|=|q_I|$ for arbitrary $I=1,..,N_f$. This operator has the largest charge among the mesonic operators $Q_IQ_J$ and has its scaling dimension-to-charge ratio at IR by
\begin{align}
    \begin{split}
        \left.\frac{\Delta^2}{q_B^2}\right|_{Q_{I_0}^2}=\frac{9}{4}\left.\left(1-\frac{N-2}{N_f}\right)^2\right/q_0^2\,.
    \end{split}
\end{align}
One can check this operator satisfies the modified Weak Gravity Conjecture
\begin{align}
    \left.\frac{\Delta^2}{q_B^2}\right|_{Q_{I_0}^2}=\frac{9}{4}\left(\frac{1}{q_0^2}\right)\left(1-\frac{N-2}{N_f}\right)^2<\frac{9}{4}\left(\frac{N_f}{\sum_{i=1}^{N_f}q_I^2}\right)\left(\frac{1}{2}\left(1-\frac{(N-2)^2}{N_f^2}\right)\right)\,,
\end{align}
where the RHS is the modified WGC bound
\begin{align}
 \frac{12(3c-2a)}{k_B}=\frac{9}{8}N_f\left.\left(1-\frac{(N-2)^2}{N_f^2}\right)\right/\left(\sum_{i=1}^{N_f} q_i^2\right)\,.
\end{align}
The mesonic operator is strictly below the modified WGC bound because of the upper bound of the conformal window $N_f<3(N-2)$.

The same argument works in an SQCD with $Sp(N)$ gauge group and $N_f$ fundamental matters in the conformal window
\begin{align}
\frac{3}{2}(N+1) < N_f< 3(N+1)\,. 
\label{eq:Spconwin}
\end{align}
In $Sp(N)$ SQCD, there is no $(Q_{I_0})^2$ since the mesonic operators $\Omega_{ab}Q_I^aQ_J^b$ are anti-symmetric in switching flavor indices $I,J$. Similarly, one can show that the mesonic operator with the largest charge does satisfy the refined WGC \eqref{eq:CFTWGC2} for every $Sp(N)$ SQCD in the conformal window \eqref{eq:Spconwin}.\footnote{One can find that the $Sp(N)$ SQCD apparently violates the refined WGC when $N=1,\,N_f=3$. However, such theory does not flow to an IR SCFT as it gets superpotential from the instanton effect.}

\subsection{$SU(N)$ gauge theory with 1 $\sym$ + 1 $\overline{\sym}$ + 2 $\antisym$ + 2 $\overline{\antisym}$ + $N_f$ $\left(\,\fund\,+\,\overline{\fund}\,\right)$}
\label{sec:nearly_conformal}
We present another counterexample of the NN-WGC. This theory is a four-dimensional $\mathcal{N}=1$ gauge theory including $N_f$ fundamental, one rank-2 symmetric, two rank-2 anti-symmetric chiral multiplets, with the chiral multiplets in their conjugate representations.

The one-loop $\beta$-function coefficient
\begin{align}
\beta\sim \sum_{i}T(\mathbf{R}_i)-3N=2\times\frac{N+2}{2}+4\times\frac{N-2}{2}+2N_f\times \frac{1}{2}-3N=N_f-2
\end{align}
is of $O(1)$ in the large $N$ limit of this theory, and the conformal window for this theory is $0\leq N_f<2$. We expect the theory to be weakly coupled at the IR fixed point, similar to the Banks-Zaks fixed point \cite{Caswell:1974gg,Banks:1981nn}.\footnote{The two-loop $\b$-function coefficient for a four-dimensional $\CN=1$ gauge theory \cite{Jones:1974pg} is given by\begin{align}
    -\frac{1}{(16\pi^2)^2} \left( 6T(\textbf{Adj})^2-2\sum_\chi T(\textbf{Adj})T(\textbf{R}_\chi)-4\sum_\chi C_2(\textbf{R}_\chi)T(\textbf{R}_\chi) \right),
\end{align}where $C_2(\textbf{R})$ denotes the quadratic Casimir for the representation \textbf{R}. At the fixed point, the coupling constant square is given as the ratio of the one-loop to the two-loop $\b$-function coefficient. We observe that this theory has a fixed point at $g^2 \sim 1/N$, featuring a weakly coupled SCFT for large $N$.} Therefore, the dual AdS cannot be in the semi-classical regime.
Meanwhile, the scaling dimension of the spectrum is almost integer-quantized since the theory is nearly conformal.
Interestingly, the theory with $N_f=0$ satisfies the NN-WGC, while the theory with $N_f=1$ does not. Here, we specifically focus on the $N_f=1$ case for our analysis.

{\renewcommand\arraystretch{1.6}
\begin{table}[h]
    \centering
    \begin{tabular}{|c|c||c|c|c|c|c|c|c|c|}
    \hline
    \# & Fields & $SU(N)$ & $ U(1)_1$& $U(1)_2$ & $U(1)_3$ & $U(1)_4$ & $U(1)_5$   & $U(1)_R$ \\\hline
    
     1 & $Q$ & $\small{\fund}$ & $1$   & 0 & 0 & 0 & $-2(N-2)$ & $R_Q$\\
     
     1 & $\widetilde{Q}$ & $\small{\overline{\fund}}$ & $-1$ & 0 & 0 & 0 & $-2(N-2)$ & $R_{{Q}}$ \\
     
     1 & $S$ & $\small{\sym}$  & 0 &1 & 0 & 1 & 0 &  $R_S$ \\
     
     1 & $\widetilde{S}$&$\small{\overline{\sym}}$ &0 & -1 & 0 & 1 & 0 &  $R_{{S}}$\\
     
     2 & $A$&$\small{\antisym}$  & 0 &0 & 1  &$-\frac{N+2}{2(N-2)}$ & 1 &  $R_A$ \\
     
     2 & $\widetilde{A}$& ${\small\overline{\antisym}}$ &  0 & 0 & -1 & $-\frac{N+2}{2(N-2)}$ & 1 & $R_{{A}}$\\\hline
    \end{tabular}
    \caption{The matter contents and their corresponding charges \label{chiral}}
\end{table}}
The matter contents and their flavor charges are given in Table \ref{chiral}. 
This theory has multiple $U(1)$ flavor charges, which require us to apply the convex hull condition discussed in Section \ref{sec:WGC}. To verify the convex hull condition, we need to find the minimum distance from the origin to the convex hull. 
It is practically useful to find an `orthonormal basis' of flavor symmetries to find the convex hull. To do this, we perform the Gram-Schmidt orthogonalization to facilitate distance calculations using the standard metric. The inner product of $U(1)$ symmetries $U(1)_i$ and $U(1)_j$ is given by the equation:
\begin{align}
    \Tr RF_iF_j=\sum_\chi (R_\chi-1)q_i(\chi)q_j(\chi)\left|\mathbf{R}_\chi\right|\ ,
\end{align}
where $\chi$, $R_\chi$, $\left|\mathbf{R_\chi}\right|$, and $q_i(\chi)$ represent the chiral multiplets, $R$-charges, dimensions of the representations $\mathbf{R}_\chi$, and $U(1)_i$ charges, respectively. If $\Tr RF_iF_j=0$, then $U(1)_i$ and $U(1)_j$ are orthogonal. The $U(1)_4$ and $U(1)_5$ are not orthogonal, so we obtain the orthogonalized $U(1)_5$ symmetry, denoted as $U(1)_{5'}$, by subtracting a fraction of $U(1)_4$ from $U(1)_5$:
\begin{align}
    U(1)_{5'}=U(1)_5-\frac{\Tr RF_4F_5}{\Tr RF_4F_4}U(1)_4\ .
\end{align}

We then proceed to normalize the orthogonalized charge-to-dimension ratios with respect to those of extremal black holes, resulting in:
\begin{align}\begin{split}
    \left.\frac{\Vec{q}}{\Delta}\right|_{\text{NN-WGC}}&=\left(\sqrt{\frac{12c}{k_{F_1}}}\frac{q_1}{\Delta},\,\dots,\,\sqrt{\frac{12c}{k_{F_{5'}}}}\frac{q_{5'}}{\Delta}\right)\ ,\\
    \left.\frac{\Vec{q}}{\Delta}\right|_{\text{modified WGC}}&=\left(\sqrt{\frac{12(3c-2a)}{k_{F_1}}}\frac{q_1}{\Delta},\,\dots,\,\sqrt{\frac{12(3c-2a)}{k_{F_{5'}}}}\frac{q_{5'}}{\Delta}\right)\ .
    \end{split}
\end{align}
The resulting normalized charge-to-dimension ratios of extremal black holes form a five-dimensional unit ball. To verify whether the WGC holds, we construct a convex hull using the set of normalized charge-to-dimension ratios of gauge-invariant operators. If this convex hull contains the five-dimensional unit ball, the WGC is satisfied. In other words, the minimum distance from the origin to the convex hull must be greater than 1.

\begin{figure}[b]
    \centering
     \begin{subfigure}[b]{0.45\textwidth}
         \includegraphics[width=\linewidth]{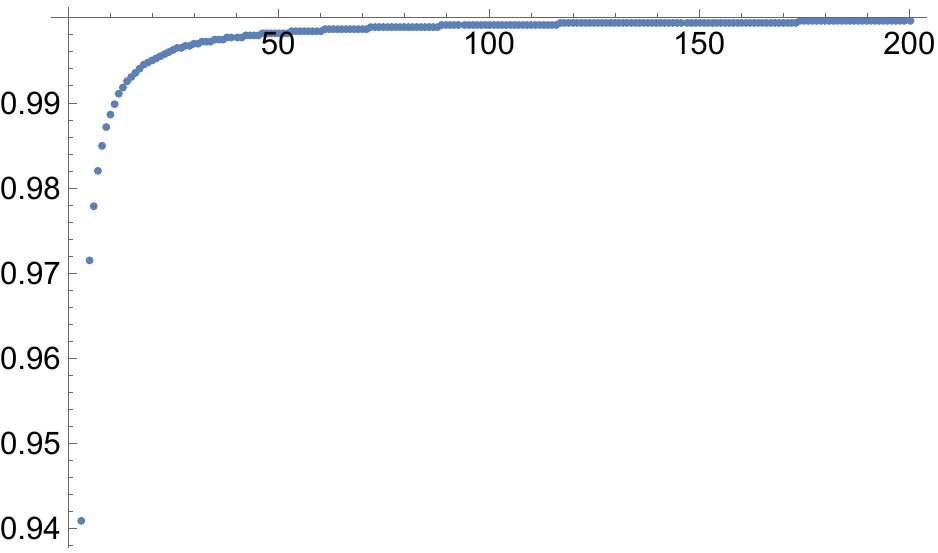}
         \caption{$a/c$ vs. $N$}
     \end{subfigure}
     \hspace{4mm}
     \begin{subfigure}[b]{0.45\textwidth}
         \includegraphics[width=\linewidth]{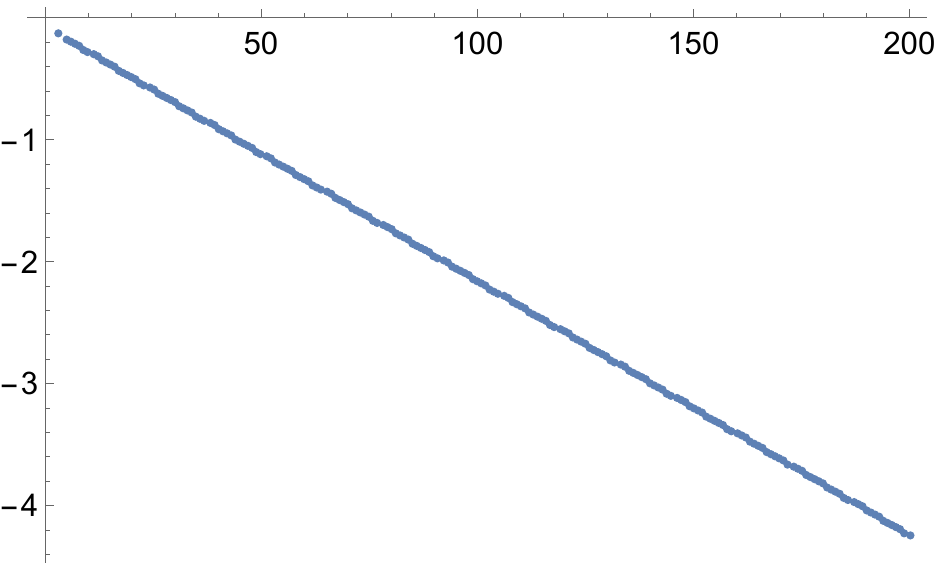}
         \caption{$a-c$ vs. $N$}
     \end{subfigure}
     \hfill
     \caption{The central charges of $SU(N)$ theory with 1$\,\sym\,$+1$\,\overline{\sym}\,$+2$\,\antisym\,$+2$\,\overline{\antisym}\,$+1$\,\left(\,\fund\,+\,\overline{\fund}\,\right)$}
     \label{fig:central_s1S1a2A2nf0} 
\end{figure}

We focus on the chiral ring operators and determine their IR $R$-charges. To do this, we impose the anomaly-free condition and use the $a$-maximization procedure \cite{Intriligator:2003jj} to find the superconformal $R$-charges at IR. Figure \ref{fig:central_s1S1a2A2nf0} illustrates the ratio and the difference of central charges $a$ and $c$ obtained through the $a$-maximization procedure. Notably, the central charge $c$ is larger than $a$, and they become equal in the large $N$ limit. This observation indicates that the normalized charge-to-dimension ratios in the modified WGC are larger (so the condition is weaker) than those in the NN-WGC.

We consider the minimal set of gauge-invariant operators that make up composite operators.
The property of chiral ring operators ensures that two operators have the same charge-to-dimension ratio if they are composed of the same ratio of matter fields. For instance, the charge-to-dimension ratio of a single trace operator $Q(S\widetilde{S})\widetilde{Q}$ is the same as that of the operator $(Q\widetilde{Q})(\Tr S\widetilde{S})$, which lies between the ratios of operators ${Q}\widetilde{{Q}}$ and $\text{Tr}(S\widetilde{S})$. Therefore, we can significantly reduce the number of operators required to construct the convex hull.

\begin{figure}[b]
    \centering
     \begin{subfigure}[b]{0.45\textwidth}
         \includegraphics[width=\linewidth]{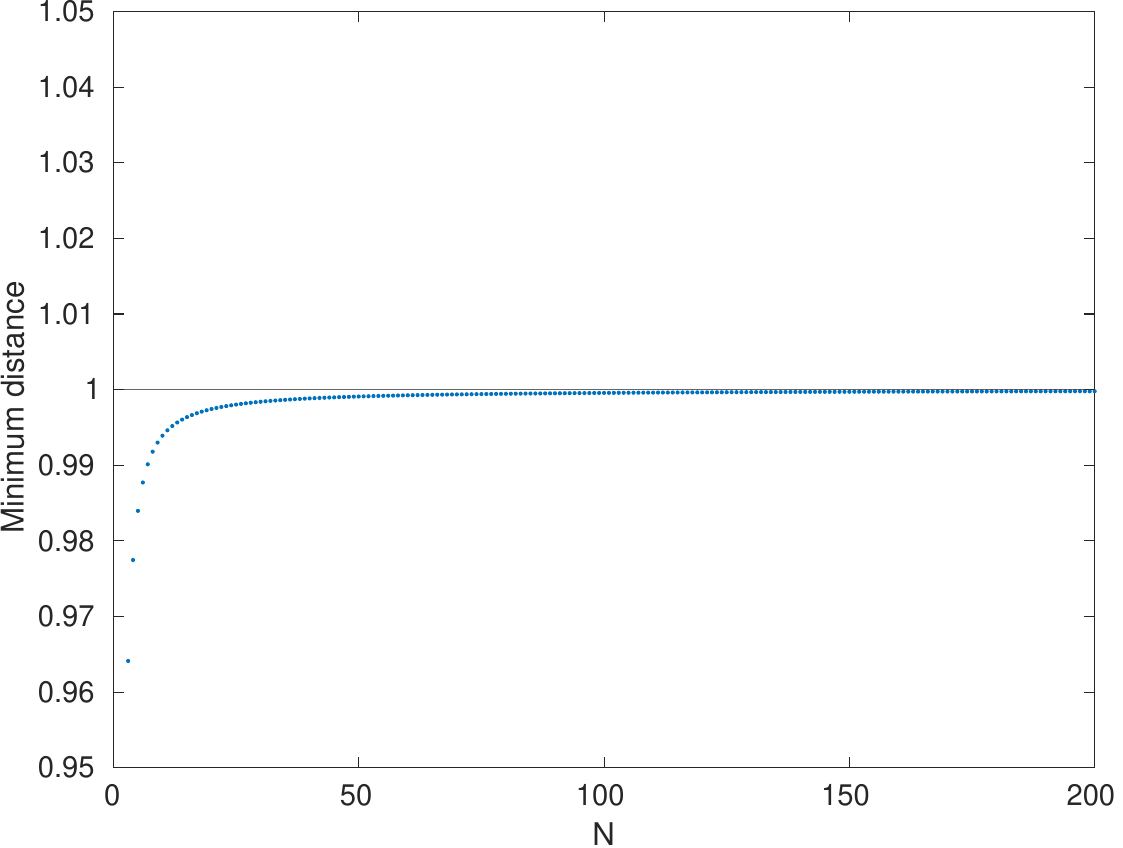}
         \caption{NN-WGC}
     \end{subfigure}
     \hspace{4mm}
     \begin{subfigure}[b]{0.45\textwidth}
         \includegraphics[width=\linewidth]{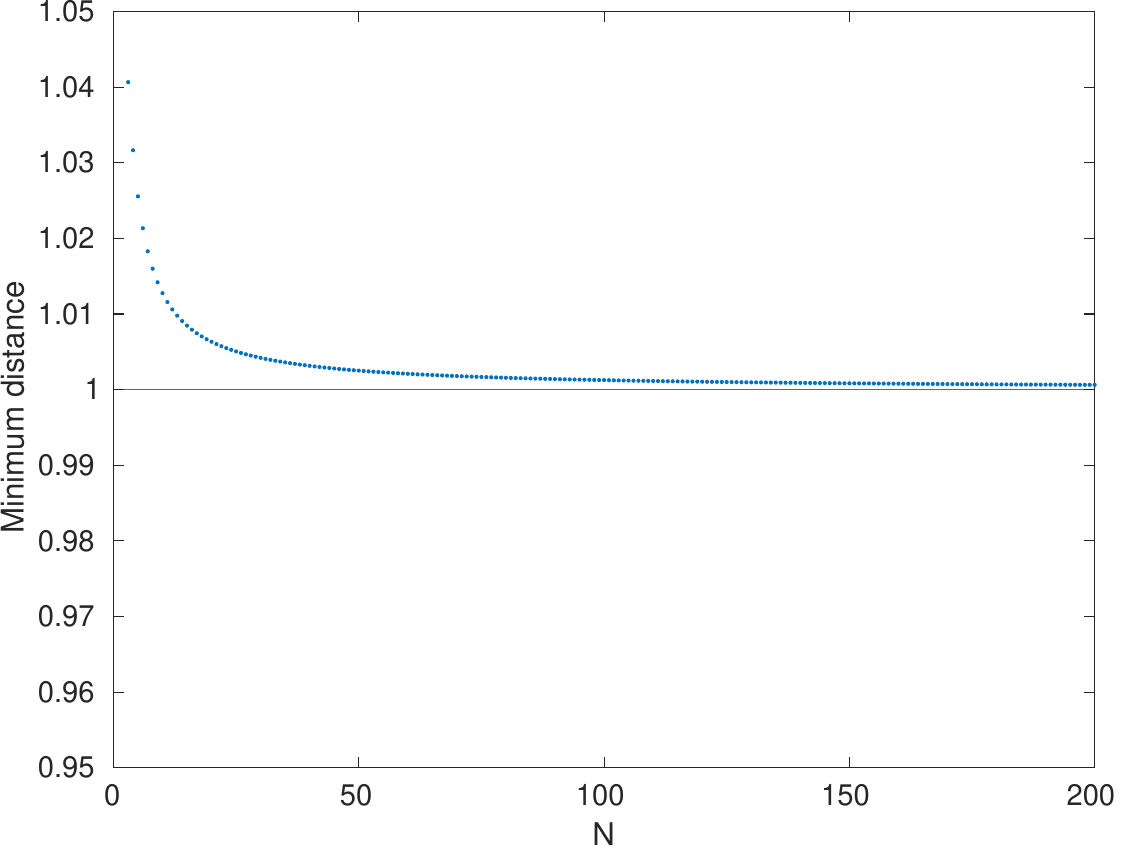}
         \caption{modified WGC}
     \end{subfigure}
     \hfill
    \caption{The minimum distance from the origin to the convex hull for the $SU(N)$ gauge theory with 1 $\sym$ + 1 $\overline{\sym}$ + 2 $\antisym$ + 2 $\overline{\antisym}$ + 1 $\,\left(\,\fund\,+\,\overline{\fund}\,\right)$}
    \label{fig:wgc_s1S1a2A2nf1}
 \end{figure}

We consider the following gauge-invariant operators (schematically):
\begin{enumerate}
    \item $\Tr S\widetilde{S},\quad\Tr A\widetilde{A},\quad \Tr (\widetilde{S}A)^2.$
    \item $Q\widetilde{Q},\quad Q\widetilde{S}Q,\quad Q\widetilde{A}S\widetilde{S}Q,\quad Q\widetilde{A}A\widetilde{A}Q,\quad Q\widetilde{A}A\widetilde{S}Q,\quad Q\widetilde{A}S\widetilde{A}Q$.
    \item $\epsilon A^{N/2},\quad$ if $N$ is even.
    \item $\epsilon {A}^{\lfloor N/2\rfloor}Q,\quad\epsilon {A}^{\lfloor N/2\rfloor}(A\widetilde{Q})\quad$ if $N$ is odd.
    \item $\epsilon\epsilon A^{2n} S^{N-2n-m} Q^{2m},\qquad m=0,1$.
    \item The operators with conjugate representations to the above-listed operators.
\end{enumerate}
Here $\epsilon$ denotes the completely anti-symmetric tensor of $SU(N)$ and we have omitted the color and flavor indices. 
It is not necessary to consider the baryonic operators of the form $\epsilon \mathcal{Q}^{n_1}_{I_1}\dots\mathcal{Q}^{n_N}_{I_N}$, where the dressed quarks are defined as $\mathcal{Q}^{n}_{I}=(S\widetilde{S})^nQ_I$. These baryonic operators are much heavier than the operators listed above, and therefore their charge-to-dimension ratios lie within the convex hull formed by the operators mentioned above. 

Finally, we plot the minimum distance from the origin to the convex hull for the NN-WGC and the modified WGC against varying $N$ in Figure \ref{fig:wgc_s1S1a2A2nf1}. This theory does not satisfy the NN-WGC for any $N$, while the modified WGC is satisfied for all $N$.

\subsection{$SU(N)$ gauge theory with 2 $\sym$ + 2 $\overline{\sym}$ + $N_f$ $\left(\,\fund\,+\,\overline{\fund}\,\right)$}
\label{sec:2sym}
In this subsection, we analyze a four-dimensional $\CN=1$ $SU(N)$ gauge theory with two rank-2 symmetric, $N_f$ fundamental chiral multiplets and their corresponding chiral multiplets in the conjugate representations. The one-loop $\beta$-function coefficient is given by
\begin{align}
\beta\sim \sum_{i}T(\mathbf{R}_i)-3N=4\times\frac{N+2}{2}+2N_f\times\frac{1}{2}-3N=N_f-N+4\ ,
\end{align}
and the conformal window for this theory is $0\leq N_f < N-4$ \cite{Agarwal:2020pol}.

The matter fields and their flavor charges are listed in Table \ref{s2S2}. 
We focus on the following gauge-invariant operators, which are enough to make a convex hull:
\begin{enumerate}
    \item $\text{Tr}Q\widetilde{Q}$,\quad$\text{Tr}S\widetilde{S}$.
    \item $Q\widetilde{S}Q,\quad \widetilde{Q}S\widetilde{Q}$.
    \item $\epsilon\epsilon S^{N-n} Q^{2n},\quad \epsilon\epsilon \widetilde{S}^{N-n}\widetilde{Q}^{2n}, \qquad n\leq N_f$.
\end{enumerate}
Again, we omitted the gauge and flavor indices.

{\renewcommand\arraystretch{1.4}
    \begin{table}[h]
        \centering
        \begin{tabular}{|c|c||c|c|c|c|c|}
        \hline
        \# & Fields & $SU(N)$ & $U(1)_S$ & $U(1)_A$ & $ U(1)_B$ & $U(1)_R$ \\\hline
         $ N_f$ & $Q$ & $\fund$ & 0  & $-\frac{2(N+2)}{N_f}$& 1 & $R_Q$  \\ 
         $ N_f$ & $\widetilde{Q}$ & $\overline{\fund}$ & 0 &  $-\frac{2(N+2)}{N_f}$ &-1 & $R_Q$ \\
         2 &$S$ & $\sym$ & 1&  1 &0 & $R_S$ \\
         2 & $\widetilde{S}$ & $\overline{\sym}$ & -1 &  1 &0 & $R_S$  \\\hline
        \end{tabular}
        \caption{The matter contents and their corresponding charges \label{s2S2}}
    \end{table}}

\begin{figure}[h]
    \centering
     \begin{subfigure}[b]{0.45\textwidth}
     \centering
    \includegraphics[width=\linewidth]{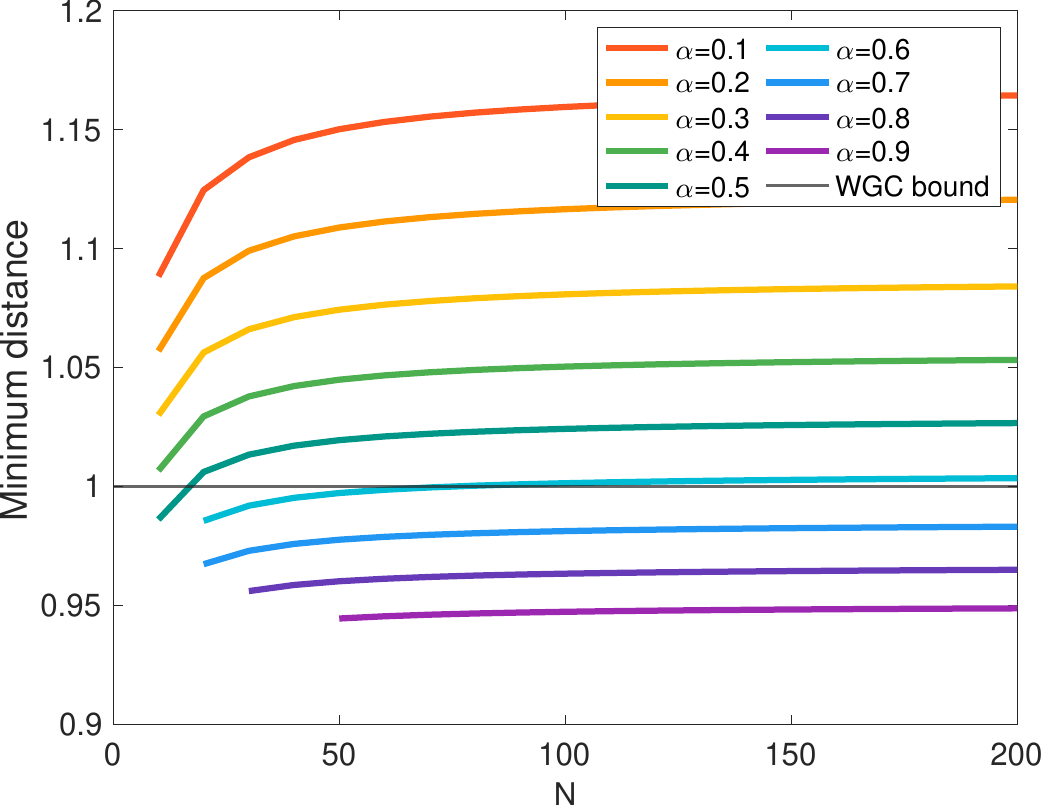}
    \caption{NN-WGC}
     \end{subfigure}
     \hspace{4mm}
     \begin{subfigure}[b]{0.45\textwidth}
     \centering
        \includegraphics[width=\linewidth]{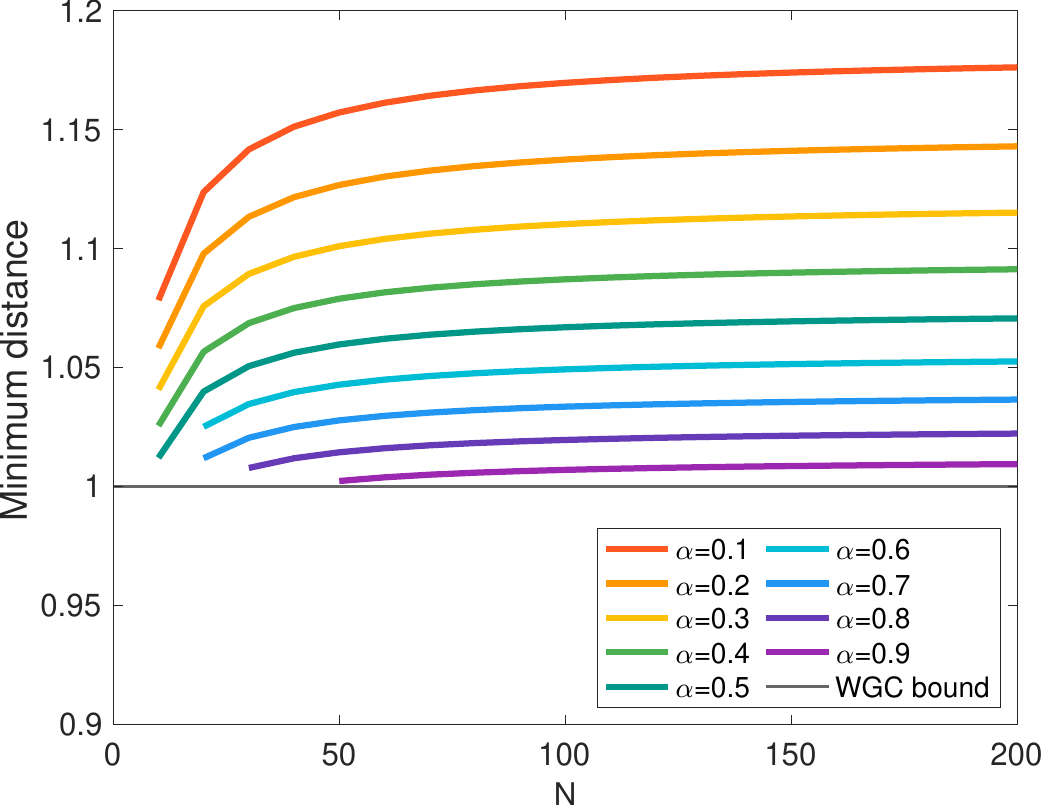}
        \caption{modified WGC}
     \end{subfigure}
     \hfill
        \caption{WGC for $SU(N)$ gauge theory with 2 $\sym$ + 2 $\overline{\sym}$ + $\alpha N$ $\left(\,\fund\,+\,\overline{\fund}\,\right)$}
        \label{fig:wgc_s2S2ven1}
 \end{figure}

Let us consider this theory in the Veneziano-like limit, where we take the ratio $\a=N_f/N$ to be a fixed value. Figure \ref{fig:wgc_s2S2ven1} shows that some theories with large values of $\alpha$ do not satisfy the NN-WGC for all possible values of $N$. We see that when $\alpha$ is small ($\alpha \lesssim 0.6$), the NN-WGC is satisfied for large enough $N$. However, when $\alpha$ is large ($\alpha \gtrsim 0.6$), the NN-WGC is not satisfied for any value of $N$. This is a priori not surprising since the Veneziano-like limit is not necessarily holographic (since the rank of the bulk gauge group increases with $N$). However, it is worth emphasizing again that the modified WGC holds even for these non-holographic theories (for arbitrary $\alpha$ and $N$). 

Let us point out that at least in the chiral ring, it is impossible to find an operator that satisfies the NN-WGC in this theory when $\alpha \gtrsim 0.6\,$. If we construct a convex hull using the charge-to-dimension ratios of elementary fields, $Q$, $\widetilde{Q}$, $S$, and $\widetilde{S}$, then all chiral ring gauge-invariant operators would be included within this convex hull. When we test the NN-WGC with this convex hull, the theory still does not satisfy the convex hull condition. This means that the NN-WGC can never be satisfied in this theory by concatenating these elementary chiral superfields. This does not fully rule out the possibility of the light-charged non-BPS operators satisfying the NN-WGC bound.

\subsection{\texorpdfstring{$SO(N)$ gauge theory with 1 $\sym$ + 1 $\antisym$ + $N_f$ $\fund$}{}}
The last example is a four-dimensional $\CN=1$ $SO(N)$ gauge theory with one symmetric, one anti-symmetric, and $N_f$ fundamental (vector) chiral multiplets.
The one-loop $\beta$-function coefficient is given as 
\begin{align}
\beta \sim \sum_{i}T(\mathbf{R}_i)-3(N-2)=(N+2)+(N-2)+N_f-3(N-2)=N_f-N+6 \ ,
\end{align}
and the conformal window of the theory is given by $0\leq N_f\leq N-6$ \cite{Razamat:2020pra, Agarwal:2020pol}.

{\renewcommand\arraystretch{1.6}
\begin{table}[h]
    \centering
    \begin{tabular}{|c|c||c|c|c|c|}
    \hline
    \# & Fields & $SO(N)$ &$U(1)_1$& $U(1)_2$ & $U(1)_R$ \\\hline

    $N_f$ & $Q$ & $\fund$ &  $N+2$& $0$  & $R_Q$\\
    
     1 & $S$ &$\sym$ & $-N_f$  & $N-2$  &$R_S$ \\
     
     1 & $A$ &$\antisym$& 0 & $-(N+2)$  &$R_A$\\\hline
    \end{tabular}
    \caption{The matter contents and their corresponding charges \label{tab:s1a1}}
\end{table}}
The matter contents and their $U(1)$ flavor charges are listed in Table \ref{tab:s1a1}.
It is enough to consider the gauge-invariant operators of the form $\Tr S^2$, $\Tr A^2$, and $Q^2$ to test the WGC. These operators have the same charge-to-dimension ratios as matter fields $Q$, $A$, and $S$ so that the charge-to-dimension ratios of the remaining chiral ring operators lie within the convex hull formed by these three operators. 

\begin{figure}[b]
    \centering
     \begin{subfigure}[b]{0.45\textwidth}
     \centering
    \includegraphics[width=\linewidth]{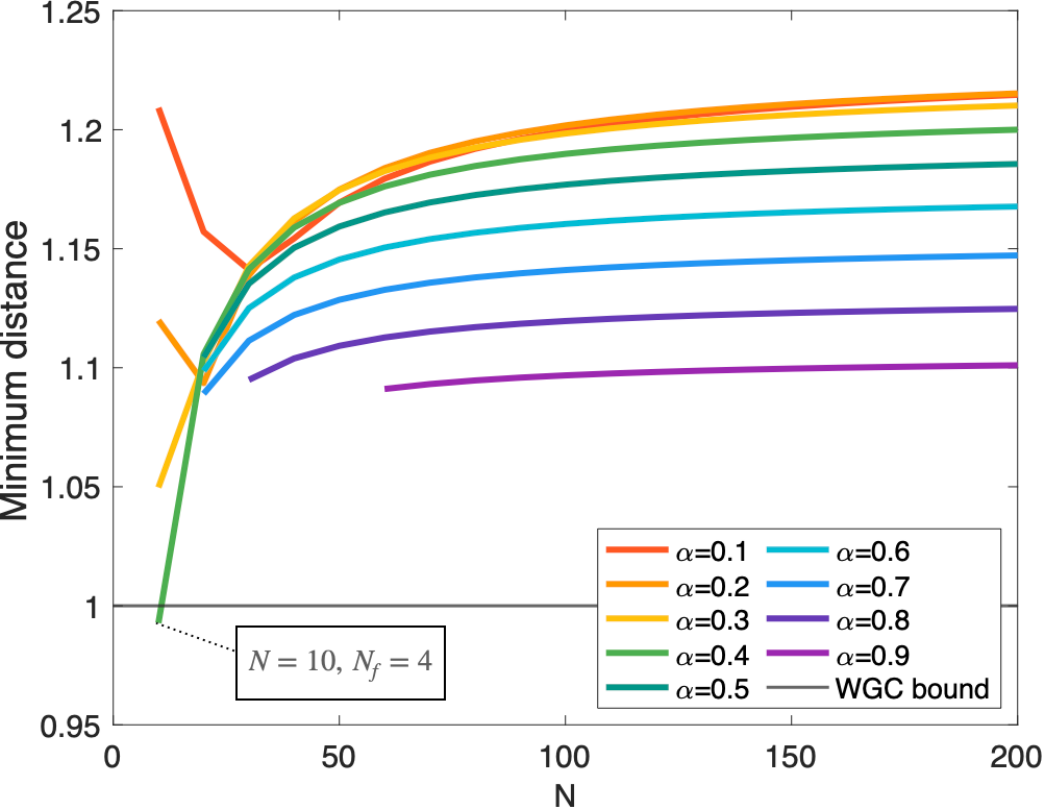}
    \caption{NN-WGC}
     \end{subfigure}
     \hspace{4mm}
     \begin{subfigure}[b]{0.45\textwidth}
     \centering
        \includegraphics[width=\linewidth]{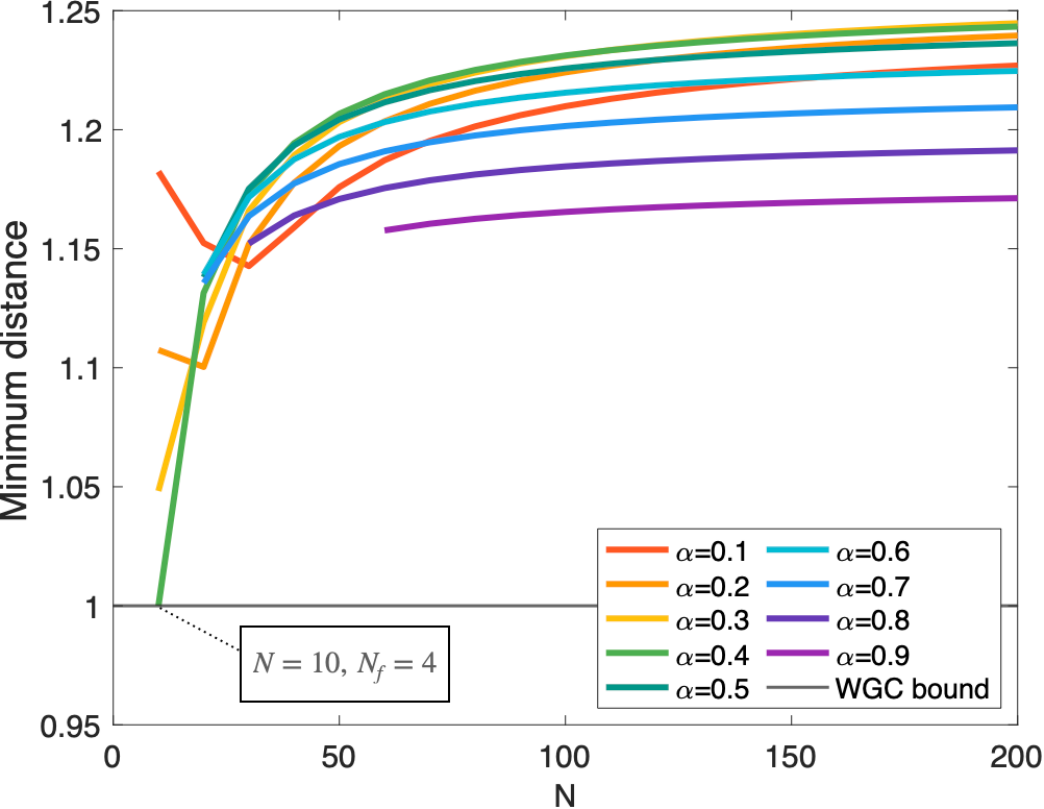}
        \caption{modified WGC}
     \end{subfigure}
     \hfill
        \caption{WGC for $SO(N)$ gauge theory with 1 $\sym$ + 1 $\antisym$ + $\a N$ $\fund$}
        \label{fig:wgc_s1a1ven}
 \end{figure}
 
In the Veneziano-like limit, the minimum distance from the origin to the convex hull is depicted in Figure \ref{fig:wgc_s1a1ven}. We observe that the NN-WGC is violated when $N=10$ and $N_f=4$, as indicated by the minimum distance being smaller than 1. For this case, the modified WGC is still satisfied, and curiously we find that the minimum distance becomes exactly 1. For all the other choices of $N$ and $N_f$, we do not find the case when the NN-WGC is violated. 

Now, let us focus on this particular case with $N=10$ and $N_f=4$. 
This theory has a vanishing 1-loop beta function and has a conformal manifold \cite{Razamat:2020pra}. 
Let us define orthogonalized flavor symmetries as
\begin{align}
    U(1)'_1=\frac{1}{4}U(1)_1 \ ,\quad U(1)'_2=-\frac{3}{2}U(1)_1-\frac{23}{4}U(1)_2 \ .
\end{align}
The $R$-charges of matter fields and the central charges are given by
\begin{align}
    R_Q=R_S=R_A=\frac{2}{3} \ ,\quad a=\frac{34}{3} \ , \quad c=\frac{137}{12} \ , 
\end{align}
and the flavor central charges are given by
\begin{align}
    k_{F_1}=-3\Tr RF_1F_1=414 \ ,\qquad k_{F_{2}}=-3\Tr RF_2F_2=313605 \ .
\end{align}

\begin{figure}[b]
    \centering
     \begin{subfigure}[b]{0.35\textwidth}
     \centering
    \includegraphics[width=\linewidth]{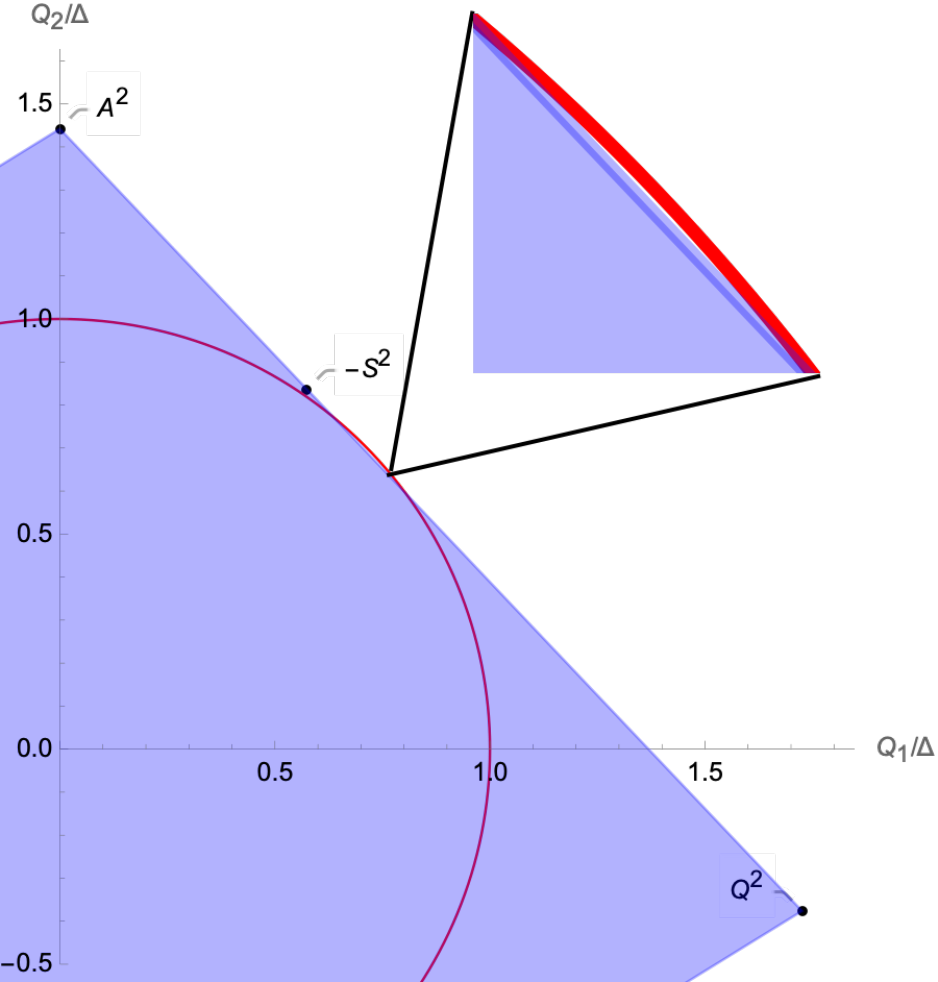}
    \caption{NN-WGC}
     \end{subfigure}
     \hspace{4mm}
     \begin{subfigure}[b]{0.35\textwidth}
     \centering
        \includegraphics[width=\linewidth]{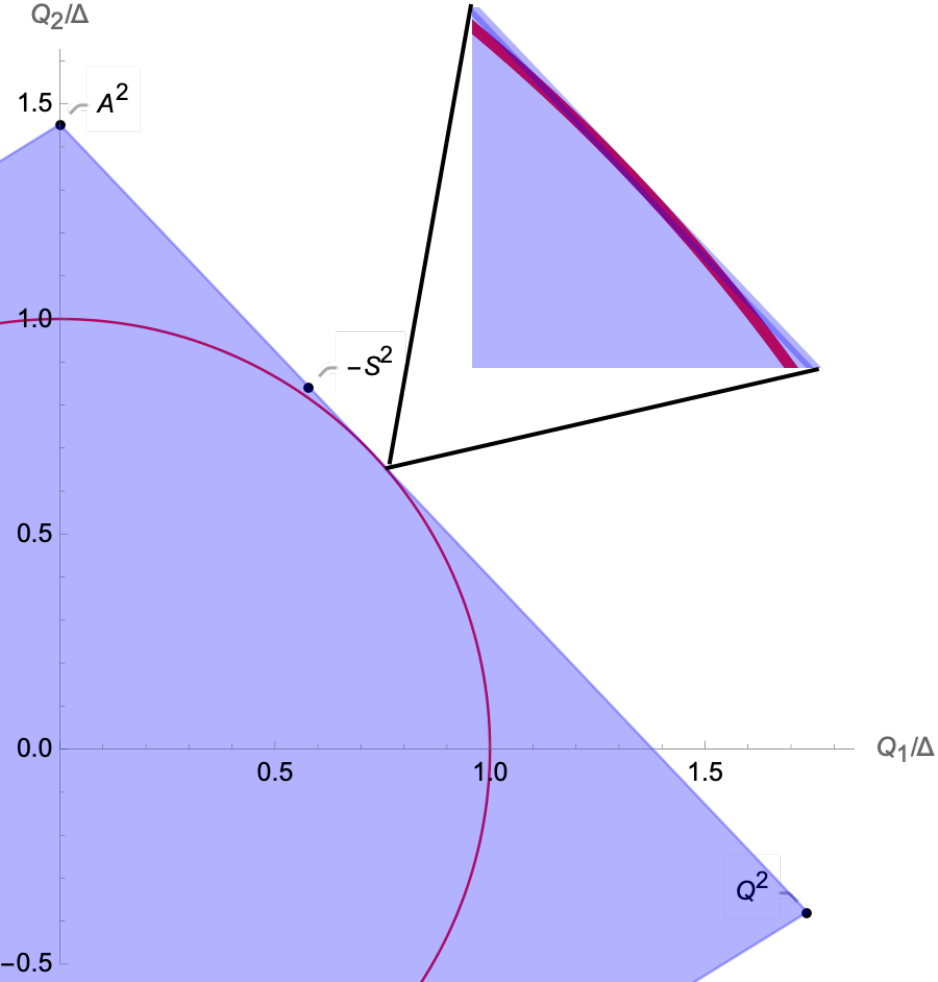}
        \caption{modified WGC}
     \end{subfigure}
     \hfill
        \caption{The convex hull formed by normalized charge-to-dimension ratios of gauge-invariant operators in the $SO(10)$ gauge theory with 1 $\sym$ + 1 $\antisym$ + $4$ $\fund$}
        \label{fig:wgc_s1a1}
 \end{figure}

Now the normalized charge-to-dimension ratios of operators are given as
\begin{align}
\begin{split}
    \left.\frac{\vec{q}}{\D}\right|_{\Tr S^2}=\left(\sqrt{\frac{12c}{k_{F_1}}}\frac{q_1}{\D},\sqrt{\frac{12c}{k_{F_2}}}\frac{q_2}{\D}\right)_{\Tr S^2}=\left(-\frac{1}{3}\sqrt{\frac{137}{46}} , -\frac{8}{3}\sqrt{\frac{685}{6969}} \right),\\
    \left.\frac{\vec{q}}{\D}\right|_{\Tr A^2}=\left(0,\sqrt{\frac{3151}{1515}}\right),\qquad
    \left.\frac{\vec{q}}{\D}\right|_{ Q^2}=\left(\sqrt{\frac{137}{46}},-2\sqrt{\frac{411}{11615}}\right) , 
\end{split}
\end{align}
for the NN-WGC. 
The convex hull formed by these charge-to-dimension ratios is visualized on the left side in Figure \ref{fig:wgc_s1a1}. The minimum distance from the origin to the convex hull is the distance to the line connecting the charge-to-dimension ratios of operators $\Tr A^2$ and $Q^2$. The minimum distance is calculated to be $\sqrt{137/139}$, which is slightly smaller than 1. Hence, the NN-WGC is not satisfied with this theory.

On the other hand, for the modified WGC, the normalized charge-to-dimension ratios of operators are given by
\begin{align}
\begin{split}
    \left.\frac{\vec{q}}{\D}\right|_{\Tr S^2}=\left(\sqrt{\frac{12(3c-2a)}{k_{F_1}}}\frac{q_1}{\D},\sqrt{\frac{12(3c-2a)}{k_{F_2}}}\frac{q_2}{\D}\right)_{\Tr S^2}=\left(-\frac{1}{3}\sqrt{\frac{139}{46}} , -\frac{8}{3}\sqrt{\frac{695}{6969}} \right),\\
    \left.\frac{\vec{q}}{\D}\right|_{\Tr A^2}=\left(0,\sqrt{\frac{3197}{1515}}\right),\qquad
    \left.\frac{\vec{q}}{\D}\right|_{ Q^2}=\left(\sqrt{\frac{139}{46}},-2\sqrt{\frac{417}{11615}}\right).
\end{split}
\end{align}
The convex hull formed by the above charge-to-dimension ratios is depicted on the RHS in Figure \ref{fig:wgc_s1a1}. Similar to the previous case, the minimum distance is the distance to the line connecting the charge-to-dimension ratios of operators $\Tr A^2$ and $Q^2$. Remarkably, the minimum distance is exactly 1 in this case, which precisely saturates the WGC bound.

\section{Large N superconformal gauge theories}
\label{sec:large_N_gauge}
In this section, we test our WGC against a large set of superconformal gauge theories. Namely, we consider four-dimensional (Lagrangian) superconformal gauge theories that admit 
a suitable large $N$ limit \cite{Agarwal:2020pol, Classification}. We focus on gauge theories with simple gauge groups that are asymptotically free and flow to a non-trivial IR fixed point. It turns out that such theories can be classified, and we find that our WGC holds for all such theories. The detail will be reported elsewhere \cite{Classification}, and here we give the summary of the result. 

The one-loop $\beta$-function coefficient for the gauge coupling in a 4d supersymmetric gauge theory with gauge group $G$ is proportional to $b_0 =  3h^{\vee}(G)-\sum_{\chi}T(\bf{R}_\chi) $, where the sum is over the chiral superfields $\chi$ in the representation $R_\chi$ of $G$. 
The condition $b_0>0$ (asymptotic freedom), along with the large $N$ limit, constrains the representation of the allowed matter fields. Especially, higher-rank (greater than two) tensor representation is not allowed since its Dynkin index increases faster than $h^{\vee}\sim N$ (here $N=\textrm{rank}(G)$). Therefore, we are only allowed to have rank-one or rank-two tensors of $G$, which greatly simplifies our classification. For simplification, we further assume that there is no superpotential.\footnote{Adding a superpotential and also flip fields can often dramatically change the IR physics \cite{Barnes:2004jj, Maruyoshi:2018nod}. We leave this analysis as future work.}

Once we have the list of viable gauge theories, the strategy is the same as in section \ref{sec:ex}:
\begin{enumerate}
\item Identify anomaly-free global symmetries. 
\item Find the superconformal $R$-charge at the IR fixed point using $a$-maximization.
\item Set an orthonormal basis of flavor $U(1)$'s using the Gram-Schmidt method so that
\begin{align}
    \text{Tr}RF_iF_j=\delta_{i,j}\,.
\end{align}
\item List the gauge-invariant chiral operators. In practice, it suffices to consider only the operators, whose charge-to-dimension ratios are extreme, that form the vertices of the convex hull.
\item Compute the distance from the origin to the convex hull.
\item Repeat the process for different $N$ and $N_f$.
\end{enumerate}
The detailed results will appear in our upcoming work \cite{Classification}, and we summarize our results for the $SU(N)$ gauge theories regarding the WGC in table \ref{tab:SUlist}. 
We find that the NN-WGC is not universally satisfied, but only under certain conditions. 
On the other hand, our modified WGC is always satisfied independent of the parameters in the theory. Therefore, at least among the superconformal gauge theories, the (modified) WGC universally holds even if the CFT is not holographically dual to a weakly coupled Einstein gravity. This result strongly supports our version of the WGC. Moreover, we find that the light-charged states, which we find to satisfy the WGC, are all in the BPS sector.

{\renewcommand\arraystretch{1.6}
\begin{table}[H]
	\centering
	\begin{tabular}{|c|c|c|}
	\hline
		Theory &  NN-WGC & Conformal window  \\\hline \hline 
        $N_f$  ( {$\fund$} + {$\overline{\fund}$} ) & $\alpha \lesssim 2.1$  & $\frac{3}{2}N < N_f<3N$ \\\hline
        1 \,\textbf{Adj}\, + $N_f$  ( $\fund$ + {$\overline{\fund}$} )& Always  & $1 \leq N_f< 2N$\\ \hline
        1 $\sym$\, + 1 $\overline{\sym}$\, + $N_f$  ( {$\fund$} + {$\overline{\fund}$} ) & $\alpha\lesssim 1.3$  & $0\le N_f < 2N-2$\\\hline
        1 $\antisym$\, + 1 $\overline{\antisym}$\, + $N_f$   ( {$\fund$} + {$\overline{\fund}$} ) & $\alpha\lesssim 0.3$  & $4\le N_f< 2N+2$ \\ \hline
        1 $\sym$\, + 1 $\overline{\antisym}$\, + 8 $\overline{\fund}$\, + $N_f$  ( {$\fund$} + {$\overline{\fund}$} ) & $\alpha\lesssim 1.3$ & $0\le N_f \leq 2N-4^*$  {\rule[-2ex]{0pt}{-3.0ex}} \\ \hline
	2 $\sym$\, + 2 $\overline{\sym}$\, + $N_f$  ( {$\fund$} + {$\overline{\fund}$} ) & $\alpha\lesssim 0.6$ & $0\le N_f< N-4$  \\ \hline
	1 $\sym$\, + 2 $\overline{\sym}$\, + 1 $\antisym$\, + 8 $\fund$\, + $N_f$  ( {$\fund$} + {$\overline{\fund}$} ) & $\alpha \lesssim 0.6$  & $0\le N_f< N-6$ \\ \hline
	1 $\sym$\, + 1 $\overline{\sym}$\, + 1 $\antisym$\, + 1 $\overline{\antisym}$\, + $N_f$  ( {$\fund$} + {$\overline{\fund}$} )& $\alpha\lesssim 0.6$  & $0\le N_f \leq N^*$ \\\hline
        
	1 $\sym$\, + 1 $\antisym$\, + 2 $\overline{\antisym}$\, + 8 $\overline{\fund}$\, + $N_f$  ( {$\fund$} + {$\overline{\fund}$} ) & $\alpha \lesssim 0.9$  & $0\le N_f<N-2$ \\\hline
		2 $\sym$\, + 2 $\overline{\antisym}$\, + 16 $\overline{\fund}$\, + $N_f$  ( {\fund} + {$\overline{\fund}$} ) & Always  & $0\leq N_f<N-8$ \\\hline
		1 \,\textbf{Adj}\, + 1 $\sym$\, + 1 $\overline{\sym}$\, + $N_f$  ( {$\fund$} + {$\overline{\fund}$} ) & Always  & $0\le N_f< N-2$ \\\hline
		2 $\antisym$\, + 2 $\overline{\antisym}$\, + $N_f$  ( {\fund} + {$\overline{\fund}$} ) & $\alpha\lesssim 0.6$  & $0\le N_f< N+4$ \\\hline
		1 \,\textbf{Adj}\, + 1 $\sym$\, + 1 $\overline{\antisym}$\, + 8 $\overline{\fund}$\, + $N_f$  ( {$\fund$} + {$\overline{\fund}$} ) &  Always  & $0\le N_f \leq N-4^*$ \\\hline
		1 \,\textbf{Adj}\, + 1 $\antisym$\, + 1 $\overline{\antisym}$\, + $N_f$  ( {$\fund$} + {$\overline{\fund}$} ) & $\alpha\lesssim 0.6$  & $0< N_f< N+2$ \\\hline
		2 \,\textbf{Adj}\, + $N_f$  ( {$\fund$} + {$\overline{\fund}$} ) & Always  & $0\le N_f \leq N^*$ \\\hline
		1 ( \sym \,+ $\overline{\sym}$ )\, + 2 ( $\antisym$ \,+ $\overline{\antisym}$ )\, + $N_f$  ( {$\fund$} + {$\overline{\fund}$} ) & $N_f=0$  & $0 \le N_f < 2$ \\\hline
		3 $\antisym$\, + 3 $\overline{\antisym}$\, + $N_f$  ( {$\fund$} + {$\overline{\fund}$} ) & $N_f=0,1,2$  & $0 \le N_f < 6$ \\\hline
		1 \,\textbf{Adj}\, + 2 $\antisym$\, + 2 $\overline{\antisym}$\, + $N_f$  ( {$\fund$} + {$\overline{\fund}$} ) &{$\begin{aligned}[c]&(N_f,N)\neq(1,5),\\&(2,6),(3,6),(3,7) \end{aligned}$}   & $0\leq N_f<4$ \\\hline
		1 \,\textbf{Adj}\, + ( $\antisym$\, + $\overline{\antisym}$ ) \, + ( {$\sym$} + {$\overline{\sym}$} ) & Always  & $*$ \\\hline
		2 \,\textbf{Adj}\, + 1 $\antisym$\, + 1 $\overline{\antisym}$\, + $N_f$  ( {$\fund$} + {$\overline{\fund}$} ) & Always  & $0\leq N_f \leq 2^*$  \\\hline
		3 \textbf{Adj} & Always  & $*$  \\ \hline
	\end{tabular}
	\caption{List of all possible superconformal $SU(N)$ theories with large $N$ limit. Each theory satisfies the modified WGC. The second column lists the condition to satisfy the NN-WGC in the Veneziano-like limit (here $\alpha \equiv N_f/N$). The last column denotes the condition for the theory to flow to a superconformal fixed point. The entries with $*$ (if $N_f$ saturates the upper bound) do not flow but have non-trivial conformal manifolds \cite{Razamat:2020pra, Bhardwaj:2013qia}.}
	\label{tab:SUlist}
\end{table}}

\section{Discussion}\label{sec:discussion}

In this paper, we have proposed a modified Weak Gravity Conjecture that applies to every four-dimensional SCFT and its corresponding dual in AdS$_5$. It is motivated by the Cardy formula for the 4d superconformal index \cite{Choi:2018hmj,Kim:2019yrz,Cabo-Bizet:2019osg,Amariti:2019mgp,Cassani:2021fyv}, whose asymptotic free-energy is universally proportional to a combination of central charges $3c-2a$. For the holographic theories, this universal behavior accounts for the entropy of the large BPS black hole in AdS$_5$ \cite{Gutowski:2004ez, Gutowski:2004yv, Chong:2005hr, Kunduri:2006ek}. From this observation, we propose a new bound on the mass (scaling dimension)-to-charge ratio of the particle (operator) that has to be included in the AdS bulk (boundary CFT). We test the modified conjecture against a large set of 4d $\mathcal{N}=1$ superconformal field theories, including non-holographic ones.

Even though our modified WGC has been tested against a large set of 4d $\mathcal{N}=1$ SCFTs, we have not given any derivation. It would be desirable to find an argument on the gravity side that supports our claim. In the AdS gravity side, we expect that the correction to the WGC bound arises from the higher-derivative terms in 5d gauged supergravity.
Recently, there have been lots of progress in analyzing four-derivative terms in AdS$_5$ gravity \cite{Bobev:2021qxx,Bobev:2022bjm, Cassani:2022lrk}. However, such four-derivative corrections were studied only in 5d $\mathcal{N}=1$ minimal gauged supergravity, which only contains the gauge field corresponding to the $R$-symmetry in boundary SCFT. Namely, there exist only $R$-charged black holes in such a gravity theory. Thus our WCG bound, which works only for the non-$R$ flavor symmetries, is not applicable. In order to derive our modified WGC from the gravity side, one needs to study higher-derivative corrections to 5d $\mathcal{N}=1$ gauged supergravity containing vector multiplets, which is not available in the literature. It would be interesting to investigate the effect of such corrections in AdS supergravity.\footnote{In flat space, the higher-derivative corrections can alter the extremal bound in such a way that any extremal black hole can decay into smaller (non-extremal) black holes \cite{Kats:2006xp,Hamada:2018dde,Cheung:2018cwt}.} 

It would be interesting to derive our conjecture from the boundary CFT perspective. Our guide was the universal property of Cardy's formula for the supersymmetric index. However, Cardy's formula only tells us the `high-temperature' asymptotics, not necessarily the low-lying spectrum. In two dimensions, modular invariance relates the high-temperature and low-temperature physics and can be thought of as the origin of the AdS WGC. In the same vein, our conjecture may be viewed as a clue toward the higher-dimensional version of the modular invariance. Further clarification of our observation would be extremely interesting. Other swampland conjectures in CFT, such as the Charge Convexity Conjecture \cite{Aharony:2021mpc} or the CFT Distance Conjecture \cite{Perlmutter:2020buo} can be also connected to the WGC, which may give us further insights. 

Since our modified WGC is proposed based on the supersymmetric analog of Cardy's formula, our version of WGC may not hold for non-supersymmetric CFTs. For the case of supersymmetric CFTs, due to the Hofman-Maldacena bound \cite{Hofman:2008ar}, $3c-2a$ is always positive (it is zero for a free vector multiplet). For a non-supersymmetric CFT, this combination of central charges can be negative in principle since the bound for the central charges is now $\frac{1}{3} \le \frac{a}{c} \le \frac{31}{18}$ (the upper bound is saturated by the free Maxwell theory). Therefore, our modified WGC fails for non-supersymmetric CFTs simply because it becomes non-sensical whenever we have $3c-2a<0$. However, as far as the authors are aware, no known non-trivial CFT in 4d lies between $\frac{3}{2} \leq \frac{a}{c} \le \frac{31}{18}$. It would be interesting to look for such CFTs or find a way to rule them out.\footnote{Similarly, for 4d $\CN=1$ SCFTs, even though the central charge bound is $\frac{1}{2} \le \frac{a}{c} \le \frac{3}{2}$, there has been no known example with $a/c$ close to $3/2$. Even though there exist many theories with $a/c>1$, the largest known value of the ratio is $a/c \simeq 1.158$, which is realized by the $SU(2)$ theory with spin $3/2$ representation \cite{Intriligator:1994rx}. Notice that $1.158 < 5/4$, which is comfortably below the bound for $\CN=2$ SCFTs. It would be interesting to find out if there is a stricter bound on the ratio.} If there are no such CFTs without supersymmetry, the bound still makes sense, and we can hope that our bound also extrapolates to non-supersymmetric theories. 

We also expect that analogous analysis can be done in other dimensions. The Cardy-like formulae in other dimensions are studied in  \cite{Choi:2019zpz,Choi:2019dfu,Amariti:2023ygn} for 3d, 
\cite{Choi:2019miv} for 5d, and \cite{Nahmgoong:2019hko,Lee:2020rns} for 6d. Especially, the Cardy formula for 6d SCFTs is determined completely by a particular combination of anomaly coefficients $\alpha-4\beta+16\gamma$ \cite{Nahmgoong:2019hko}, from which we may be able to deduce analogous WGC bound. This combination of the anomaly coefficients is shown to be positive \cite{Yankielowicz:2017xkf}, just as in our case we have $3c-2a>0$.


\begin{acknowledgments}
We thank Junho Hong, Seok Kim, and Seungkyu Kim for helpful discussions. 
SC thanks KAIST for the hospitality during his visit, where this work was conceived.
This work is supported by a KIAS Individual Grant PG081602 at Korea Institute for Advanced Study (SC), and the National Research Foundation of Korea (NRF) Grant RS-2023-00208602 (MC, KL, JS). 
The work of JS is also supported by POSCO Science Fellowship of POSCO TJ Park Foundation.

\end{acknowledgments}

\bibliographystyle{JHEP}
\bibliography{refs}

\providecommand{\href}[2]{#2}\begingroup\raggedright\begin{thebibliography}{10}

\bibitem{Maldacena:1997re}
J.M.~Maldacena, \emph{{The Large N limit of superconformal field theories and
  supergravity}}, \href{https://doi.org/10.1023/A:1026654312961}{\emph{Adv.
  Theor. Math. Phys.} {\bfseries 2} (1998) 231}
  [\href{https://arxiv.org/abs/hep-th/9711200}{{\ttfamily hep-th/9711200}}].

\bibitem{Gubser:1998bc}
S.S.~Gubser, I.R.~Klebanov and A.M.~Polyakov, \emph{{Gauge theory correlators
  from noncritical string theory}},
  \href{https://doi.org/10.1016/S0370-2693(98)00377-3}{\emph{Phys. Lett. B}
  {\bfseries 428} (1998) 105}
  [\href{https://arxiv.org/abs/hep-th/9802109}{{\ttfamily hep-th/9802109}}].

\bibitem{Witten:1998qj}
E.~Witten, \emph{{Anti-de Sitter space and holography}},
  \href{https://doi.org/10.4310/ATMP.1998.v2.n2.a2}{\emph{Adv. Theor. Math.
  Phys.} {\bfseries 2} (1998) 253}
  [\href{https://arxiv.org/abs/hep-th/9802150}{{\ttfamily hep-th/9802150}}].

\bibitem{tHooft:1973alw}
G.~'t~Hooft, \emph{{A Planar Diagram Theory for Strong Interactions}},
  \href{https://doi.org/10.1016/0550-3213(74)90154-0}{\emph{Nucl. Phys. B}
  {\bfseries 72} (1974) 461}.

\bibitem{Heemskerk:2009pn}
I.~Heemskerk, J.~Penedones, J.~Polchinski and J.~Sully, \emph{{Holography from
  Conformal Field Theory}},
  \href{https://doi.org/10.1088/1126-6708/2009/10/079}{\emph{JHEP} {\bfseries
  10} (2009) 079} [\href{https://arxiv.org/abs/0907.0151}{{\ttfamily
  0907.0151}}].

\bibitem{El-Showk:2011yvt}
S.~El-Showk and K.~Papadodimas, \emph{{Emergent Spacetime and Holographic
  CFTs}}, \href{https://doi.org/10.1007/JHEP10(2012)106}{\emph{JHEP} {\bfseries
  10} (2012) 106} [\href{https://arxiv.org/abs/1101.4163}{{\ttfamily
  1101.4163}}].

\bibitem{Arkani-Hamed:2006emk}
N.~Arkani-Hamed, L.~Motl, A.~Nicolis and C.~Vafa, \emph{{The String landscape,
  black holes and gravity as the weakest force}},
  \href{https://doi.org/10.1088/1126-6708/2007/06/060}{\emph{JHEP} {\bfseries
  06} (2007) 060} [\href{https://arxiv.org/abs/hep-th/0601001}{{\ttfamily
  hep-th/0601001}}].

\bibitem{Harlow:2022gzl}
D.~Harlow, B.~Heidenreich, M.~Reece and T.~Rudelius, \emph{{The Weak Gravity
  Conjecture: A Review}},  \href{https://arxiv.org/abs/2201.08380}{{\ttfamily
  2201.08380}}.

\bibitem{Nakayama:2015hga}
Y.~Nakayama and Y.~Nomura, \emph{{Weak gravity conjecture in the AdS/CFT
  correspondence}},
  \href{https://doi.org/10.1103/PhysRevD.92.126006}{\emph{Phys. Rev. D}
  {\bfseries 92} (2015) 126006}
  [\href{https://arxiv.org/abs/1509.01647}{{\ttfamily 1509.01647}}].

\bibitem{Agarwal:2019crm}
P.~Agarwal and J.~Song, \emph{{Large N Gauge Theories with a Dense Spectrum and
  the Weak Gravity Conjecture}},
  \href{https://doi.org/10.1007/JHEP05(2021)124}{\emph{JHEP} {\bfseries 05}
  (2021) 124} [\href{https://arxiv.org/abs/1912.12881}{{\ttfamily
  1912.12881}}].

\bibitem{Agarwal:2020pol}
P.~Agarwal, K.-H.~Lee and J.~Song, \emph{{Classification of large N
  superconformal gauge theories with a dense spectrum}},
  \href{https://doi.org/10.1007/JHEP10(2021)049}{\emph{JHEP} {\bfseries 10}
  (2021) 049} [\href{https://arxiv.org/abs/2007.16165}{{\ttfamily
  2007.16165}}].

\bibitem{Montero:2016tif}
M.~Montero, G.~Shiu and P.~Soler, \emph{{The Weak Gravity Conjecture in three
  dimensions}}, \href{https://doi.org/10.1007/JHEP10(2016)159}{\emph{JHEP}
  {\bfseries 10} (2016) 159}
  [\href{https://arxiv.org/abs/1606.08438}{{\ttfamily 1606.08438}}].

\bibitem{Benjamin:2016fhe}
N.~Benjamin, E.~Dyer, A.L.~Fitzpatrick and S.~Kachru, \emph{{Universal Bounds
  on Charged States in 2d CFT and 3d Gravity}},
  \href{https://doi.org/10.1007/JHEP08(2016)041}{\emph{JHEP} {\bfseries 08}
  (2016) 041} [\href{https://arxiv.org/abs/1603.09745}{{\ttfamily
  1603.09745}}].

\bibitem{Bae:2018qym}
J.-B.~Bae, S.~Lee and J.~Song, \emph{{Modular Constraints on Superconformal
  Field Theories}}, \href{https://doi.org/10.1007/JHEP01(2019)209}{\emph{JHEP}
  {\bfseries 01} (2019) 209}
  [\href{https://arxiv.org/abs/1811.00976}{{\ttfamily 1811.00976}}].

\bibitem{Cardy:1986ie}
J.L.~Cardy, \emph{{Operator Content of Two-Dimensional Conformally Invariant
  Theories}}, \href{https://doi.org/10.1016/0550-3213(86)90552-3}{\emph{Nucl.
  Phys. B} {\bfseries 270} (1986) 186}.

\bibitem{Choi:2018hmj}
S.~Choi, J.~Kim, S.~Kim and J.~Nahmgoong, \emph{{Large AdS black holes from
  QFT}},  \href{https://arxiv.org/abs/1810.12067}{{\ttfamily 1810.12067}}.

\bibitem{Kim:2019yrz}
J.~Kim, S.~Kim and J.~Song, \emph{{A 4d $ \mathcal{N} $ = 1 Cardy Formula}},
  \href{https://doi.org/10.1007/JHEP01(2021)025}{\emph{JHEP} {\bfseries 01}
  (2021) 025} [\href{https://arxiv.org/abs/1904.03455}{{\ttfamily
  1904.03455}}].

\bibitem{Cabo-Bizet:2019osg}
A.~Cabo-Bizet, D.~Cassani, D.~Martelli and S.~Murthy, \emph{{The asymptotic
  growth of states of the 4d $ \mathcal{N}=1 $ superconformal index}},
  \href{https://doi.org/10.1007/JHEP08(2019)120}{\emph{JHEP} {\bfseries 08}
  (2019) 120} [\href{https://arxiv.org/abs/1904.05865}{{\ttfamily
  1904.05865}}].

\bibitem{Amariti:2019mgp}
A.~Amariti, I.~Garozzo and G.~Lo~Monaco, \emph{{Entropy function from toric
  geometry}},
  \href{https://doi.org/10.1016/j.nuclphysb.2021.115571}{\emph{Nucl. Phys. B}
  {\bfseries 973} (2021) 115571}
  [\href{https://arxiv.org/abs/1904.10009}{{\ttfamily 1904.10009}}].

\bibitem{Cassani:2021fyv}
D.~Cassani and Z.~Komargodski, \emph{{EFT and the SUSY Index on the 2nd
  Sheet}}, \href{https://doi.org/10.21468/SciPostPhys.11.1.004}{\emph{SciPost
  Phys.} {\bfseries 11} (2021) 004}
  [\href{https://arxiv.org/abs/2104.01464}{{\ttfamily 2104.01464}}].

\bibitem{Gutowski:2004ez}
J.B.~Gutowski and H.S.~Reall, \emph{{Supersymmetric AdS(5) black holes}},
  \href{https://doi.org/10.1088/1126-6708/2004/02/006}{\emph{JHEP} {\bfseries
  02} (2004) 006} [\href{https://arxiv.org/abs/hep-th/0401042}{{\ttfamily
  hep-th/0401042}}].

\bibitem{Gutowski:2004yv}
J.B.~Gutowski and H.S.~Reall, \emph{{General supersymmetric AdS(5) black
  holes}}, \href{https://doi.org/10.1088/1126-6708/2004/04/048}{\emph{JHEP}
  {\bfseries 04} (2004) 048}
  [\href{https://arxiv.org/abs/hep-th/0401129}{{\ttfamily hep-th/0401129}}].

\bibitem{Chong:2005hr}
Z.W.~Chong, M.~Cvetic, H.~Lu and C.N.~Pope, \emph{{General non-extremal
  rotating black holes in minimal five-dimensional gauged supergravity}},
  \href{https://doi.org/10.1103/PhysRevLett.95.161301}{\emph{Phys. Rev. Lett.}
  {\bfseries 95} (2005) 161301}
  [\href{https://arxiv.org/abs/hep-th/0506029}{{\ttfamily hep-th/0506029}}].

\bibitem{Kunduri:2006ek}
H.K.~Kunduri, J.~Lucietti and H.S.~Reall, \emph{{Supersymmetric multi-charge
  AdS(5) black holes}},
  \href{https://doi.org/10.1088/1126-6708/2006/04/036}{\emph{JHEP} {\bfseries
  04} (2006) 036} [\href{https://arxiv.org/abs/hep-th/0601156}{{\ttfamily
  hep-th/0601156}}].

\bibitem{Kinney:2005ej}
J.~Kinney, J.M.~Maldacena, S.~Minwalla and S.~Raju, \emph{{An Index for 4
  dimensional super conformal theories}},
  \href{https://doi.org/10.1007/s00220-007-0258-7}{\emph{Commun. Math. Phys.}
  {\bfseries 275} (2007) 209}
  [\href{https://arxiv.org/abs/hep-th/0510251}{{\ttfamily hep-th/0510251}}].

\bibitem{Romelsberger:2005eg}
C.~Romelsberger, \emph{{Counting chiral primaries in N = 1, d=4 superconformal
  field theories}},
  \href{https://doi.org/10.1016/j.nuclphysb.2006.03.037}{\emph{Nucl. Phys. B}
  {\bfseries 747} (2006) 329}
  [\href{https://arxiv.org/abs/hep-th/0510060}{{\ttfamily hep-th/0510060}}].

\bibitem{Hofman:2008ar}
D.M.~Hofman and J.~Maldacena, \emph{{Conformal collider physics: Energy and
  charge correlations}},
  \href{https://doi.org/10.1088/1126-6708/2008/05/012}{\emph{JHEP} {\bfseries
  05} (2008) 012} [\href{https://arxiv.org/abs/0803.1467}{{\ttfamily
  0803.1467}}].

\bibitem{Benini:2018ywd}
F.~Benini and E.~Milan, \emph{{Black Holes in 4D $\mathcal{N}$=4
  Super-Yang-Mills Field Theory}},
  \href{https://doi.org/10.1103/PhysRevX.10.021037}{\emph{Phys. Rev. X}
  {\bfseries 10} (2020) 021037}
  [\href{https://arxiv.org/abs/1812.09613}{{\ttfamily 1812.09613}}].

\bibitem{Choi:2021rxi}
S.~Choi, S.~Jeong, S.~Kim and E.~Lee, \emph{{Exact QFT duals of AdS black
  holes}},  \href{https://arxiv.org/abs/2111.10720}{{\ttfamily 2111.10720}}.

\bibitem{Bobev:2021qxx}
N.~Bobev, K.~Hristov and V.~Reys, \emph{{AdS$_{5}$ holography and
  higher-derivative supergravity}},
  \href{https://doi.org/10.1007/JHEP04(2022)088}{\emph{JHEP} {\bfseries 04}
  (2022) 088} [\href{https://arxiv.org/abs/2112.06961}{{\ttfamily
  2112.06961}}].

\bibitem{Bobev:2022bjm}
N.~Bobev, V.~Dimitrov, V.~Reys and A.~Vekemans, \emph{{Higher derivative
  corrections and AdS5 black holes}},
  \href{https://doi.org/10.1103/PhysRevD.106.L121903}{\emph{Phys. Rev. D}
  {\bfseries 106} (2022) L121903}
  [\href{https://arxiv.org/abs/2207.10671}{{\ttfamily 2207.10671}}].

\bibitem{Cassani:2022lrk}
D.~Cassani, A.~Ruip\'erez and E.~Turetta, \emph{{Corrections to AdS$_{5}$ black
  hole thermodynamics from higher-derivative supergravity}},
  \href{https://doi.org/10.1007/JHEP11(2022)059}{\emph{JHEP} {\bfseries 11}
  (2022) 059} [\href{https://arxiv.org/abs/2208.01007}{{\ttfamily
  2208.01007}}].

\bibitem{Kats:2006xp}
Y.~Kats, L.~Motl and M.~Padi, \emph{{Higher-order corrections to mass-charge
  relation of extremal black holes}},
  \href{https://doi.org/10.1088/1126-6708/2007/12/068}{\emph{JHEP} {\bfseries
  12} (2007) 068} [\href{https://arxiv.org/abs/hep-th/0606100}{{\ttfamily
  hep-th/0606100}}].

\bibitem{Cheung:2018cwt}
C.~Cheung, J.~Liu and G.N.~Remmen, \emph{{Proof of the Weak Gravity Conjecture
  from Black Hole Entropy}},
  \href{https://doi.org/10.1007/JHEP10(2018)004}{\emph{JHEP} {\bfseries 10}
  (2018) 004} [\href{https://arxiv.org/abs/1801.08546}{{\ttfamily
  1801.08546}}].

\bibitem{Hamada:2018dde}
Y.~Hamada, T.~Noumi and G.~Shiu, \emph{{Weak Gravity Conjecture from Unitarity
  and Causality}},
  \href{https://doi.org/10.1103/PhysRevLett.123.051601}{\emph{Phys. Rev. Lett.}
  {\bfseries 123} (2019) 051601}
  [\href{https://arxiv.org/abs/1810.03637}{{\ttfamily 1810.03637}}].

\bibitem{Cheung:2014vva}
C.~Cheung and G.N.~Remmen, \emph{{Naturalness and the Weak Gravity
  Conjecture}},
  \href{https://doi.org/10.1103/PhysRevLett.113.051601}{\emph{Phys. Rev. Lett.}
  {\bfseries 113} (2014) 051601}
  [\href{https://arxiv.org/abs/1402.2287}{{\ttfamily 1402.2287}}].

\bibitem{Seiberg:1994pq}
N.~Seiberg, \emph{{Electric - magnetic duality in supersymmetric nonAbelian
  gauge theories}},
  \href{https://doi.org/10.1016/0550-3213(94)00023-8}{\emph{Nucl. Phys. B}
  {\bfseries 435} (1995) 129}
  [\href{https://arxiv.org/abs/hep-th/9411149}{{\ttfamily hep-th/9411149}}].

\bibitem{Anselmi:1997am}
D.~Anselmi, D.Z.~Freedman, M.T.~Grisaru and A.A.~Johansen,
  \emph{{Nonperturbative formulas for central functions of supersymmetric gauge
  theories}}, \href{https://doi.org/10.1016/S0550-3213(98)00278-8}{\emph{Nucl.
  Phys. B} {\bfseries 526} (1998) 543}
  [\href{https://arxiv.org/abs/hep-th/9708042}{{\ttfamily hep-th/9708042}}].

\bibitem{Intriligator:1995id}
K.A.~Intriligator and N.~Seiberg, \emph{{Duality, monopoles, dyons, confinement
  and oblique confinement in supersymmetric SO(N(c)) gauge theories}},
  \href{https://doi.org/10.1016/0550-3213(95)00159-P}{\emph{Nucl. Phys. B}
  {\bfseries 444} (1995) 125}
  [\href{https://arxiv.org/abs/hep-th/9503179}{{\ttfamily hep-th/9503179}}].

\bibitem{Intriligator:2003jj}
K.A.~Intriligator and B.~Wecht, \emph{{The Exact superconformal R symmetry
  maximizes a}},
  \href{https://doi.org/10.1016/S0550-3213(03)00459-0}{\emph{Nucl. Phys. B}
  {\bfseries 667} (2003) 183}
  [\href{https://arxiv.org/abs/hep-th/0304128}{{\ttfamily hep-th/0304128}}].

\bibitem{Caswell:1974gg}
W.E.~Caswell, \emph{{Asymptotic Behavior of Nonabelian Gauge Theories to Two
  Loop Order}}, \href{https://doi.org/10.1103/PhysRevLett.33.244}{\emph{Phys.
  Rev. Lett.} {\bfseries 33} (1974) 244}.

\bibitem{Banks:1981nn}
T.~Banks and A.~Zaks, \emph{{On the Phase Structure of Vector-Like Gauge
  Theories with Massless Fermions}},
  \href{https://doi.org/10.1016/0550-3213(82)90035-9}{\emph{Nucl. Phys. B}
  {\bfseries 196} (1982) 189}.

\bibitem{Jones:1974pg}
D.R.T.~Jones, \emph{{Asymptotic Behavior of Supersymmetric Yang-Mills Theories
  in the Two Loop Approximation}},
  \href{https://doi.org/10.1016/0550-3213(75)90256-4}{\emph{Nucl. Phys. B}
  {\bfseries 87} (1975) 127}.

\bibitem{Razamat:2020pra}
S.S.~Razamat, E.~Sabag and G.~Zafrir, \emph{{Weakly coupled conformal manifolds
  in 4d}}, \href{https://doi.org/10.1007/JHEP06(2020)179}{\emph{JHEP}
  {\bfseries 06} (2020) 179}
  [\href{https://arxiv.org/abs/2004.07097}{{\ttfamily 2004.07097}}].

\bibitem{Classification}
M.~Cho, K.-H.~Lee and J.~Song, \emph{{Classification of Large N Superconformal
  Gauge Theories in 4d}}, {\emph{To appear} }.

\bibitem{Barnes:2004jj}
E.~Barnes, K.A.~Intriligator, B.~Wecht and J.~Wright, \emph{{Evidence for the
  strongest version of the 4d a-theorem, via a-maximization along RG flows}},
  \href{https://doi.org/10.1016/j.nuclphysb.2004.09.016}{\emph{Nucl. Phys. B}
  {\bfseries 702} (2004) 131}
  [\href{https://arxiv.org/abs/hep-th/0408156}{{\ttfamily hep-th/0408156}}].

\bibitem{Maruyoshi:2018nod}
K.~Maruyoshi, E.~Nardoni and J.~Song, \emph{{Landscape of Simple Superconformal
  Field Theories in 4d}},
  \href{https://doi.org/10.1103/PhysRevLett.122.121601}{\emph{Phys. Rev. Lett.}
  {\bfseries 122} (2019) 121601}
  [\href{https://arxiv.org/abs/1806.08353}{{\ttfamily 1806.08353}}].

\bibitem{Bhardwaj:2013qia}
L.~Bhardwaj and Y.~Tachikawa, \emph{{Classification of 4d N=2 gauge theories}},
  \href{https://doi.org/10.1007/JHEP12(2013)100}{\emph{JHEP} {\bfseries 12}
  (2013) 100} [\href{https://arxiv.org/abs/1309.5160}{{\ttfamily 1309.5160}}].

\bibitem{Aharony:2021mpc}
O.~Aharony and E.~Palti, \emph{{Convexity of charged operators in CFTs and the
  weak gravity conjecture}},
  \href{https://doi.org/10.1103/PhysRevD.104.126005}{\emph{Phys. Rev. D}
  {\bfseries 104} (2021) 126005}
  [\href{https://arxiv.org/abs/2108.04594}{{\ttfamily 2108.04594}}].

\bibitem{Perlmutter:2020buo}
E.~Perlmutter, L.~Rastelli, C.~Vafa and I.~Valenzuela, \emph{{A CFT distance
  conjecture}}, \href{https://doi.org/10.1007/JHEP10(2021)070}{\emph{JHEP}
  {\bfseries 10} (2021) 070}
  [\href{https://arxiv.org/abs/2011.10040}{{\ttfamily 2011.10040}}].

\bibitem{Intriligator:1994rx}
K.A.~Intriligator, N.~Seiberg and S.H.~Shenker, \emph{{Proposal for a simple
  model of dynamical SUSY breaking}},
  \href{https://doi.org/10.1016/0370-2693(94)01336-B}{\emph{Phys. Lett. B}
  {\bfseries 342} (1995) 152}
  [\href{https://arxiv.org/abs/hep-ph/9410203}{{\ttfamily hep-ph/9410203}}].

\bibitem{Choi:2019zpz}
S.~Choi, C.~Hwang and S.~Kim, \emph{{Quantum vortices, M2-branes and black
  holes}},  \href{https://arxiv.org/abs/1908.02470}{{\ttfamily 1908.02470}}.

\bibitem{Choi:2019dfu}
S.~Choi and C.~Hwang, \emph{{Universal 3d Cardy Block and Black Hole Entropy}},
  \href{https://doi.org/10.1007/JHEP03(2020)068}{\emph{JHEP} {\bfseries 03}
  (2020) 068} [\href{https://arxiv.org/abs/1911.01448}{{\ttfamily
  1911.01448}}].

\bibitem{Amariti:2023ygn}
A.~Amariti, J.~Nian, L.A.~Pando~Zayas and A.~Segati, \emph{{Universal
  Cardy-Like Behavior of 3D A-Twisted Partition Functions}},
  \href{https://arxiv.org/abs/2306.05462}{{\ttfamily 2306.05462}}.

\bibitem{Choi:2019miv}
S.~Choi and S.~Kim, \emph{{Large AdS$_6$ black holes from CFT$_5$}},
  \href{https://arxiv.org/abs/1904.01164}{{\ttfamily 1904.01164}}.

\bibitem{Nahmgoong:2019hko}
J.~Nahmgoong, \emph{{6d superconformal Cardy formulas}},
  \href{https://doi.org/10.1007/JHEP02(2021)092}{\emph{JHEP} {\bfseries 02}
  (2021) 092} [\href{https://arxiv.org/abs/1907.12582}{{\ttfamily
  1907.12582}}].

\bibitem{Lee:2020rns}
K.~Lee and J.~Nahmgoong, \emph{{Cardy Limits of 6d Superconformal Theories}},
  \href{https://doi.org/10.1007/JHEP05(2021)118}{\emph{JHEP} {\bfseries 05}
  (2021) 118} [\href{https://arxiv.org/abs/2006.10294}{{\ttfamily
  2006.10294}}].

\bibitem{Yankielowicz:2017xkf}
S.~Yankielowicz and Y.~Zhou, \emph{{Supersymmetric R\'enyi entropy and
  Anomalies in 6d (1,0) SCFTs}},
  \href{https://doi.org/10.1007/JHEP04(2017)128}{\emph{JHEP} {\bfseries 04}
  (2017) 128} [\href{https://arxiv.org/abs/1702.03518}{{\ttfamily
  1702.03518}}].

\end{thebibliography}\endgroup

\end{document}